\documentclass{emulateapj}

\usepackage{natbib}
\usepackage{hyperref}
\shorttitle{Molecules and MHSE}
\shortauthors{Jaeggli, Lin, \& Uitenbroek}

\newcommand{\hh}{H$_2$}

\begin{document}

\title{On Molecular Hydrogen Formation and the Magnetohydrostatic Equilibrium of Sunspots}

\author{S.~A. Jaeggli\altaffilmark{1} and H. Lin}
\affil{Institute for Astronomy, University of Hawai'i, 2680 Woodlawn Drive, Honolulu, HI 96822, USA}
\and
\author{H. Uitenbroek}
\affil{National Solar Observatory, Sacramento Peak, P.O. Box 62, Sunspot, NM 88349, USA}

\altaffiltext{1}{Visiting Student, National Solar Observatory, operated by the Association of Universities for Research in Astronomy, Inc. (AURA), under cooperative agreement with the National Science Foundation.}

\begin{abstract}
We have investigated the problem of sunspot magnetohydrostatic equilibrium with comprehensive IR sunspot magnetic field survey observations of the highly sensitive \ion{Fe}{1} lines at 15650 \AA\ and nearby OH lines.  We have found that some sunspots show isothermal increases in umbral magnetic field strength which cannot be explained by the simplified sunspot model with a single-component ideal gas atmosphere assumed in previous investigations.  Large sunspots universally display non-linear increases in magnetic pressure over temperature, while small sunspots and pores display linear behavior.  The formation of molecules provides a mechanism for isothermal concentration of the umbral magnetic field, and we propose that this may explain the observed rapid increase in umbral magnetic field strength relative to temperature.  Existing multi-component sunspot atmospheric models predict that a significant amount of molecular hydrogen (\hh) exists in the sunspot umbra.  The formation of \hh\ can significantly alter the thermodynamic properties of the sunspot atmosphere and may play a significant role in sunspot evolution.  In addition to the survey observations, we have performed detailed chemical equilibrium calculations with full consideration of radiative transfer effects to establish OH as a proxy for \hh, and demonstrate that a significant population of \hh\ exists in the coolest regions of large sunspots.
\end{abstract}

\keywords{magnetic fields --- molecular processes --- Sun: infrared --- Sun: sunspots}


\section{Introduction}
\subsection{Background}
Sunspot umbrae are quasi-stable structures that can be considered to be in nearly magnetohydrostatic (MHS) equilibrium due to their long lifetimes relative to the dynamical timescale of the quiet sun \citep{meyer77}.  Hale's historic observation of the Zeeman effect in sunspots revealed them as concentrations of strong magnetic fields \citep{hale08}, and Biermann and Alfv\'en provided us with the first physically plausible explanation of the sunspot phenomenon \citep{biermann41,alfven43}.  Although the sunspot atmosphere is cool and has a low degree of ionization, its conductivity is high, and it can be assumed that the magnetic field is ``frozen in'' to the gas.  In Alfv\'en's theory the strong magnetic field suppresses the convective heating of the interior of the sunspot. It also supports the cool sunspot interior against the higher pressure of the hotter surrounding quiet sun \citep{deinzer65}.  For a sunspot with circular symmetry the MHS equilibrium state at any radius from the center of the sunspot ($r$) and height ($z$) in the sunspot atmosphere can be written as:
\begin{equation}
P_{qs}(z) = n_s(r,z)kT_s(r,z) + {1 \over 8 \pi}(B_z^2(r,z) + F_c(r,z)),
\label{eqn1}
\end{equation}
\noindent for c.g.s. units, where $P_{qs}$ is the gas pressure in the quiet sun, and the pressure in the sunspot is given by the product of the number density of the gas ($n_s$), Boltzmann's constant ($k$), and temperature ($T_s$).  The magnetic field contributes a pressure term from the vertical component of the field ($B_z$) and a tension or curvature term ($F_c$) which is given by:
\begin{equation}
F_c(r,z)= 2\int_r^a B_z(r^\prime,z) \ {\partial B_r(r^\prime,z) \over \partial z} \ dr^\prime,
\label{eqn2}
\end{equation}
\noindent where the integral runs from the radius in question to the maximum radius of the sunspot ($a$), and requires knowledge of the vertical gradient of the radial component of the magnetic field ($B_r$) \citep{cowling76,martinez93}.

Although this theory provides us with a basic description of the MHS equilibrium condition for sunspots, observational tests are difficult to carry out, even for the form of circular symmetry assumed in Equation \ref{eqn1}.  Because of radiative transfer (RT) effects, the magnetic and thermodynamic properties of a sunspot observed in regions of different temperature do not originate at the same geometrical depth in the sunspot atmosphere, and information about the vertical structure cannot be derived easily. Therefore the contribution of the curvature force term, which requires knowledge of the vertical gradient of the magnetic field, cannot be determined directly from observations.  Nevertheless, under the restrictive conditions present in the sunspot umbra it is possible to make some simplifying assumptions.  Here the magnetic fields are mostly vertical and the contribution from the curvature force term is expected to be small.  In addition, the radial gradient of temperature is small, and the magnetic field and temperature observations can be considered to originate at a single height.  Therefore, if we make the assumptions that sunspots are vertical magnetic flux tubes with constant magnetic field strength, and have an ideal gas atmosphere with constant density and height, then Equation \ref{eqn1} can be simplified to the thermal-magnetic relation:
\begin{equation}
B^2(r) \propto T_{qs} - T_s(r).
\label{eqn3}
\end{equation}
\begin{figure}
\centerline{\includegraphics[scale=0.5]{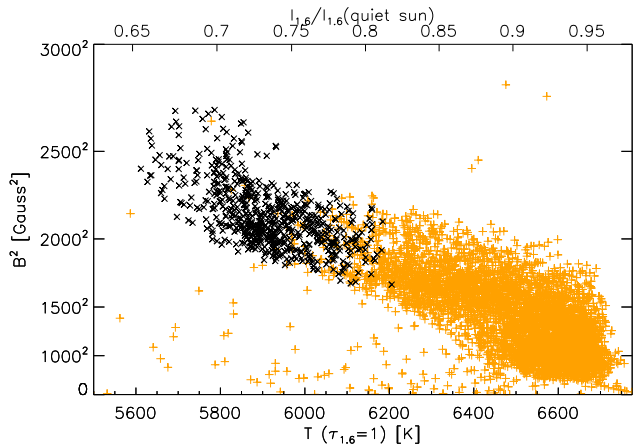}}
\caption{The $B^2$ vs $T$ relation from NOAA 11130 observed with FIRS on December 2, 2010 displaying an isothermal increase in the umbral magnetic field.  Points selected from the penumbra are shown in gold (+) and points from the umbra are black ($\times$).}
\label{fig1}
\end{figure}

Although Equation \ref{eqn3} is elegantly simple it has not been verified observationally. While some examples of the empirical relation between $B^2$ and $T$ determined from sunspots are best described by a linear relationship \citep{gurman81, livingston02, penn02, penn03b}, the majority of modern observations show that $B^2$ can be a highly non-linear function of $T$ \citep{balthasar93, kopp93, lites93, martinez93, solanki93, stanchfield97, westendorp01, penn03a, mathew04}. Figure \ref{fig1} gives an example of the thermal-magnetic relation for the main sunspot in NOAA 11130 that we have recently derived from spectropolarimetric observations at 15650 \AA, where temperature has been determined from the continuum contrast, and magnetic field was obtained by Milne-Eddington inversion of the \ion{Fe}{1} line pair. The non-linearities in the $B^2$ vs. $T$ relation are evident:  the sunspot shows an unexpected sharp increase in magnetic field strength at relatively constant temperature in the darkest part of the umbra. This result is puzzling. With the simplifying assumptions leading to Equation \ref{eqn3}, Figure \ref{fig1} implies the existence of a heretofore unidentified mechanism at work in the coolest regions that reduces the thermal pressure of the sunspot atmosphere to accommodate the higher magnetic pressure in these regions.

\subsection{The Magnetohydrostatic Equilibrium of Sunspots and the Role of Molecular Hydrogen}
In deriving Equation \ref{eqn3} we have made the assumptions that:
1) sunspot magnetic fields are near vertical with respect to the local solar surface and that the curvature force has a negligible effect on the force balance of a sunspot,
2) the geometrical depth of the $\tau=1$ level depends roughly on the temperature $T$, and that the temperature and magnetic field observed in a sunspot originates from roughly the same height in the sunspot atmosphere, and
3) the density of the solar atmosphere is a constant that does not change with temperature.
In assuming a flux tube model, we also neglect fine structure, flows, growth, decay, and other evolutionary effects.  While the first two assumptions are difficult to verify due to radiative transfer effects, they may be justified under restrictive conditions such as those encountered in the sunspot umbra. On the other hand, the third assumption of constant density is obviously not warranted, especially under the temperature regime encountered in the umbral photosphere. 

At an effective temperature of 5,750 K the quiet sun photosphere is largely neutral and already harbors many heavy-element molecular species with high dissociation energies \citep{grevesse94}. At even lower temperatures in the sunspot umbra many more molecular species, in particular molecular hydrogen (\hh), are able to form.  Because the abundances of heavy elements in the solar photosphere are very low compared to hydrogen, the number of molecules formed containing heavy elements is correspondingly low with respect to the total particle number density, therefore the formation of heavy element molecules cannot have a significant effect on the thermodynamic properties of the solar atmosphere. However, in the umbra the atmosphere is very cool, making it possible to form a substantial fraction of \hh.  Observations of the fluorescent \hh\ lines in the ultraviolet have confirmed its presence in the chromosphere above sunspots \citep{jordan78, bartoe79, innes08}, while atmospheric models of the sunspot umbra predict a molecular hydrogen population of up to $10\%$, peaking near the height of continuum formation \citep{maltby86}.

\begin{deluxetable*}{cccrrccc}
\tablecaption{Summary of Selected Observations \label{tbl1}}
\tablewidth{0pt}
\tablehead{
\colhead{Target Name} & \colhead{Date} & \colhead{UT Time\tablenotemark{a}} & \colhead{Lat [$^\circ$]} & \colhead{Lon [$^\circ$]} & \colhead{$\mu$\tablenotemark{b}} & \colhead{Classification} & \colhead{Phase}
}
\startdata
NOAA 9429$^{c}$	& 2001-04-18	& 14:44:17	& 8.39	& 13.42	& 0.947 & $\alpha$ & decay \\ 
NOAA 11035		& 2009-12-17	& 15:58:13	& 27.73	& 28.83	& 0.754 & $\beta\gamma$ & growth \\ 
NOAA 11046		& 2010-02-13	& 17:30:53	& 22.88	& 3.41	& 0.859 & $\beta$ & decay \\ 
NOAA 11049		& 2010-02-19	& 15:57:02	& -20.61	& 17.72	& 0.933 & $\beta$ & quiescent \\ 
NOAA 11101		& 2010-09-02	& 14:27:11	& 10.97	& 38.70	& 0.782 & $\alpha$ & quiescent \\ 
NOAA 11130		& 2010-12-02	& 15:42:30	& 11.03	& 49.91	& 0.624 & $\beta$ & decay \\ 
NOAA 11131		& 2010-12-06	& 17:00:35	& 29.78	& -18.45	& 0.815 & $\alpha$ & quiescent \\ 
\enddata
\tablenotetext{a}{Start time of the observation.}
\tablenotetext{b}{Cosine of the heliocentric angle.}
\tablenotetext{c}{Preliminary observation taken with the Horizontal Spectrograph.}
\end{deluxetable*}

The formation of a large fraction of molecules may have important effects on the thermodynamic properties of the solar atmosphere and the physics of sunspots.  For example, as free atoms combine into molecules the dissociation energy is released. This may increase the local thermal energy content of the atmosphere if it cannot be dissipated rapidly enough. On the other hand, molecules have the ability to store energy in rotational and vibrational degrees of freedom which do not contribute to the thermal signature of the gas, resulting in an increased heat capacity.  Therefore if a significant fraction of the hydrogen atoms in the photosphere of a sunspot exist in molecular form, then the thermodynamic properties of the sunspot will be significantly different from that of the quiet sun photosphere.  Finally, the combination of free atoms into molecules decreases the total particle number density and pressure of the gas.  This effect is of particular importance to the problem of MHS equilibrium in sunspots since it provides a mechanism for changing the gas pressure of the sunspot atmosphere without a corresponding change in the temperature, which may explain the isothermal intensification of the magnetic field strength in the darkest regions of the sunspot seen in Figure \ref{fig1}.  Even a small alteration in the balance of gas pressures in a sunspot may translate into large changes in the sunspot magnetic field.

\subsection{Objectives of This Research}
Given the critical role that \hh\ may potentially play in the physics of sunspots, it is the objective of this study to determine if a significant fraction of \hh\ exists in the photosphere of sunspots.  We will also demonstrate that \hh\ formation occurs at a temperature coinciding with the sharp rise of $B^2$ seen in Figure \ref{fig1} to support the argument that this feature is due to the formation of \hh.

\subsection{Methodology}
\label{sec_meth}
To establish the existence of \hh\ in the sunspot photosphere and its link with the isothermal intensification of the umbral magnetic fields, it is necessary to characterize the 2D magnetic field, temperature, and \hh\ fraction of many sunspots at different stages of evolution.  In previous investigations, the authors presented the thermal-magnetic relation from a single sunspot observation, or from observations of many sunspots with very limited spatial resolution.  To fully explore the relationship between temperature and magnetic field strength, we have conducted a comprehensive survey of sunspots, obtaining high spectral and spatial resolution spectropolarimetric observations of the infrared \ion{Fe}{1} line pair at 15650 \AA. The instrumentation and observational setup, as well as the dataset, are described in \S\ref{sec_obs}. The reduction of the data, including the treatment of scattered light and the measurement and calibration of the telescope and instrumental polarization crosstalk, will be described in \S\ref{sec_data_reduction} and \ref{sec_pol_cor}. We have derived the vector magnetic field configuration of the sunspots using a Milne-Eddington inversion code (\S\ref{sec_inversion}). 

Due to the highly forbidden nature of the infrared ro-vibrational transitions of \hh\ in the solar atmosphere, and the current lack of space-based instruments to provide spectral coverage of the fluorescent emission lines in the UV, direct measurement of the abundance of \hh abundance in the photosphere lies beyond our grasp. Nevertheless, since the formation of the molecules is directly controlled by the temperature of the gas, it is possible to infer the photospheric \hh\ abundance from proxy measurements of other similar molecules. The dissociation energy of the \hh\ molecule (4.48 eV) is very similar to the dissociation energy of the hydroxide (OH) molecule (4.39 eV) which exhibits strong absorption features throughout the visible and IR sunspot spectrum.  Serendipitously, it has strong absorption lines inside the spectral window of the \ion{Fe}{1} line pair at 15650 \AA\ used for the measurement of sunspot magnetic fields.  We have established the equivalent width of the OH 15652 \AA\ line as a proxy for the \hh\ abundance through detailed calculation of the chemical equilibrium, and OH spectral line synthesis for model solar atmospheres with different effective temperatures.  The details of this calculation are described in \S\ref{sec_RT_model}. 

We will present the results of this study, including the measurement of the magnetic field pressure $B^2$, the equivalent width of the OH 15652 \AA\ line, and the inferred \hh\ abundance as a function of the temperature of the sunspot atmosphere in \S\ref{sec_results}. Finally, we will summarize this study and discuss the influence of \hh\ formation on the emergence and evolution of sunspots in \S\ref{sec_conclusions}.

\section{Observations and Instrumentation}\label{sec_obs}
The data present in this paper were obtained with the Dunn Solar Telescope (DST) at the National Solar Observatory (NSO) located in Sunspot, New Mexico.  Preliminary observations were taken with the Horizontal Spectrograph (HSG) in 2001 and 2005, and the main survey observations were taken over a period from July 2009 to December 2010 using the newly constructed Facility IR Spectropolarimeter (FIRS).  For every sunspot in the survey we have made spectropolarimetric observations of a 10 \AA\ bandpass at 15650 \AA\, covering the \ion{Fe}{1} 15648.5 \AA\ g=3 and \ion{Fe}{1} 15652.9 \AA\ g=1.7 lines, a blend of two OH lines at 15650.8 \AA, and two OH lines on either side of the \ion{Fe}{1} g=1.67 line at 15651.9 and 15653.7 \AA.  The complete survey contains 66 observations of 23 different active regions and pores.  From this sample we present seven illustrative cases.  Table \ref{tbl1} summarizes the observed active regions, the date and time of observation, and their heliocentric position.  The full survey will be presented and discussed at length in \citet{jaeggliphdt}.
\begin{figure*}
\centerline{\includegraphics[scale=0.69]{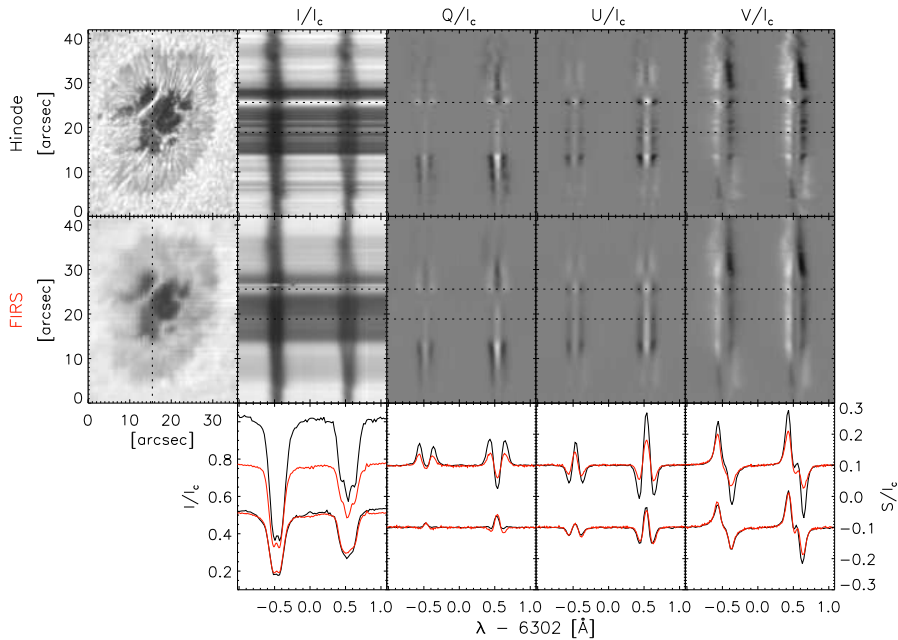}}
\caption{Comparison of data from Hinode SOT/SP and FIRS, using coincident sunspot observations taken on 2009-07-07.  Shown are the intensity maps, full Stokes spectra, and line spectra from FIRS (red) and Hinode (black) at the positions indicated by the dotted lines.}
\label{fig2}
\end{figure*}

The Horizontal Spectrograph (HSG) is a versatile and reconfigurable 3 m spectrograph arm with multiple available gratings.  It consists of an adjustable rectangular slit, collimator lens, plane grating, camera lens, and detector.  The experimental HSG setup for these sunspot observations was similar to those described in \citet{lin98} and \citet{lin95}.  During the 2001 observations a $256\times256$ HgCdTe NICMOS 3 detector was used to record Stokes spectra.

FIRS is a new facility instrument for the DST recently completed by the Institute for Astronomy at the University of Hawai'i (P.I. Haosheng Lin) in collaboration with the NSO.  Several unique design elements allow FIRS to perform full-Stokes vector spectropolarimetry at two wavelengths simultaneously in the visible and infrared, at  6302 and 10830 \AA\ or 6302 and 15650 \AA.  An off-axis near-Littrow configuration and steeply blazed grating provide high spectral resolution, and spectra are imaged using a Kodak $2048\times2048$ CCD for the visible and a Raytheon Virgo $1024\times1024$ HgCdTe array for the infrared.  FIRS maps four slit positions simultaneously to increase observing cadence and make efficient use of the large format of these detectors.  A scan area of $150 \times 75$ arcsec$^2$ can be covered in 20 minutes with 0.29 arcsec$^2$ spatial pixels, while achieving a signal-to-noise of 1000.  Specialized narrow band filters for each wavelength prevent the spectra from adjacent slits from overlapping.  Diffraction-limited performance at 6302 \AA\ can be achieved with FIRS using the DST's High Order Adaptive Optics (HOAO) system \citep{rimmele04}, which also provides excellent image correction at infrared wavelengths even in mediocre seeing conditions.  Separate liquid crystal variable retarders for each wavelength modulate the polarization in an efficiency-balanced scheme which is analyzed and split into two orthogonally polarized beams by a Wollaston prism which allows the spurious seeing effects to be eliminated in post-processing.  A more detailed description of the instrument can be found in \citet{jaeggli10} and soon in Lin et al. (in prep.).

\section{Data Reduction and Analysis}
\begin{figure*}[!ht]
\centerline{\includegraphics[scale=0.69]{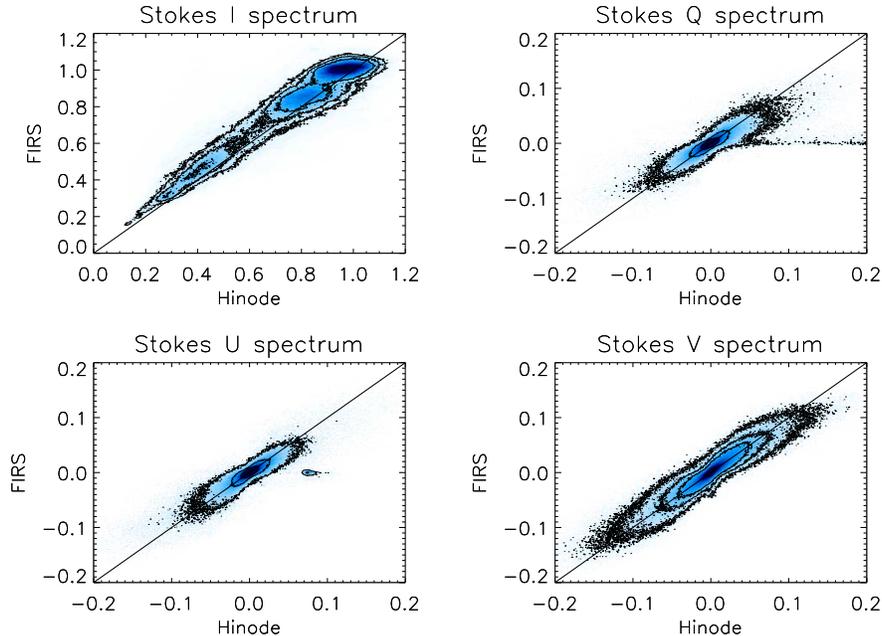}}
\caption{Point -by-point scatter plot comparison of the full Stokes spectra from Hinode SOT/SP and FIRS data cubes, using coincident sunspot observations taken on 2009-07-07.  The color scale indicates the density of points, 50, 75, 95\% contours are displayed for I and 90, 95, 99\% contours for Q, U, and V.  The solid line indicates the 1:1 correspondence.}
\label{fig3}
\end{figure*}

\subsection{Data Reduction}
\label{sec_data_reduction}
The observations from FIRS were processed using the IDL reduction tools (http://kopiko.ifa.hawaii.edu/firs/) which corrected for the infrared detector non-linearity, subtracted darks, and demodulated the raw polarization states to retrieve the Stokes vector.  A flat field was produced by fitting and removing absorption lines from spatially averaged quiet sun observations and was applied to Stokes I to remove intensity variations due to the detector, slit, narrow-band filter, and other elements in the optical system.  The demodulation scheme for the Stokes vector effectively removes response variation under the assumption that the filter profile is flat over the spectral window.  The spectra from each slit were corrected for spectral curvature and drift, the orthogonally polarized beams were combined, and the spectra were stacked into a data cube.  Following the reduction the data cubes were cropped spatially to a relevant area around each sunspot.

Observations from the HSG were reduced in a similar fashion to the FIRS observations (details in \citet{lin98} and \citet{lin95}), however no correction was applied to compensate for the detector linearity.

\subsection{Correction of Instrumental Polarization}
\label{sec_pol_cor}
The Stokes Q, U, and V spectra were corrected for instrumental polarization cross-talk using the method of \citet{kuhn94}.  This approach rather simply assumes that the Stokes Q and U profiles should be symmetric about the line center and Stokes V should be anti-symmetric.  From this assumption the cross-talk coefficients are determined from a small portion of the spectrum and applied to the full data cube.

Based on these assumptions, one would think that the Kuhn method might not work well for physical situations which produce non- symmetric and anti-symmetric Stokes profiles.  Therefore, we have verified the appropriateness of the Kuhn method, and the data quality in general, by comparing corrected FIRS observations to reduced spectra from the Hinode Solar Optical Telescope Spectropolarimeter (SOT/SP) which were closely coincident in time (about 4 minutes apart).  The polarization cross-talk in the SOT/SP has been very carefully measured and corrected as described by \citet{ichimoto08}.  The maps and example spectra from each observation are shown in Figure \ref{fig2}, and in Figure \ref{fig3} we show a point-by-point comparison of all quiet sun and sunspot spatial and spectral positions from the aligned datacubes of Stokes spectra from FIRS and the SOT/SP.  Most predominantly noticeable is the effect of seeing on the observations.  Small scale features with extreme intensity, velocity, or polarization signatures are smoothed over in the FIRS spectrum, such as the upper line spectrum in Figure \ref{fig2} which is from a position in the light bridge.  However, the overall behavior of the spectra is similar, and non-symmetric profiles seen in the Hinode data are faithfully reproduced by FIRS.  In the scatter plots the overall adherence to the 1:1 line (esp. at low signal levels in Stokes I) is encouraging.  The flattening of Stokes Q, U, and V in Figure \ref{fig3} indicates that the highest polarization signals seen by Hinode are not detected by FIRS, these signals are spatially discrete and become smoothed over by seeing.  The high Q and U values in the Hinode data where corresponding FIRS values are zero, are the result of a bad scan-step on the penumbral/quiet sun boundary in the SOT/SP scan.

\subsection{Inversion of Sunspot Magnetic Fields}
\label{sec_inversion}
We have developed a Milne-Eddington inversion code, called the Two-Component Magneto-Optical (2CMO) Inversion Code, based on the formalism of \citet{jefferies89} for the inference of the vector magnetic field configuration from our data.  This code models the observed Stokes spectra with a magnetic and a non-magnetic atmospheric component, fully accounting for magneto-optical effects.  Voigt and Faraday-Voigt functions are used for the synthesis of the \ion{Fe}{1} line profiles, while Gaussian profiles are used for fitting the OH lines.  The 2CMO inversion code is capable of simultaneously fitting of multiple spectral lines with different Land\'e $g_{eff}$ factors, such as the line pairs of the \ion{Fe}{1} 630 nm and 1565 nm lines, to achieve higher magnetic field strength sensitivity in the weak-field regime \citep{solanki92}.

The magnetic component in 2CMO is described by ten parameters:  magnetic filling fraction ($f$), magnetic field strength ($B$), inclination ($\gamma$), azimuth ($\chi$), source function ($B_0$), source function gradient ($B_1$), line center ($\lambda$), doppler width ($\Delta\lambda$), damping parameter ($a$), and the ratio of line to continuum absorption ($\eta_0$).  By assuming $\eta_0$ has the same behavior between the two lines the number of fitted parameters can be decreased.  The value of $\eta_0$ is scaled for the atomic lines by a constant factor of $g_l f e^{-E_l/{kT}}$, where $g_l$ is the degeneracy of the lower level, $f$ is the oscillator strength, $E_l$ is the energy of the lower level for the transition, and T is the temperature (we assume an average temperature at the height of line formation of 5000 K).  The line values $log(gf)$ and $g_{eff}$ for the \ion{Fe}{1} lines were taken from \citet{borrero03}.

The OH lines are very weak in regions warmer than the sunspot umbra, and are therefore susceptible to stray light in much the same way as the atomic lines.  In addition, the OH lines experience the molecular Zeeman effect \citep{berdyugina02} which produces non-negligible Stokes V signal at umbral magnetic field strengths.  In order to accurately fit the Stokes V profiles during the inversion, and properly account for the stray light contamination for the OH lines, they must be included in the magnetic component.   The OH lines are assumed to be influenced by the same magnetic field as the atomic lines, however the gaussian width and depth of the OH lines are left as independent parameters.
\begin{figure*}
\centerline{\includegraphics[scale=0.69]{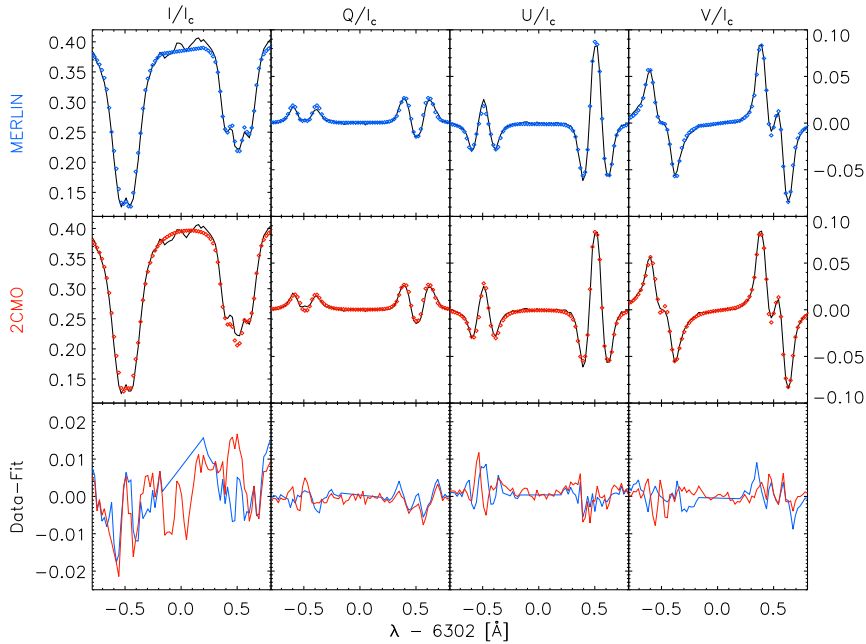}}
\caption{Fitted line profiles from the MERLIN (blue) and 2CMO (red) inversions with the original Stokes spectrum from Hinode (black points), from the umbra of NOAA 11024.}
\label{fig4}
\end{figure*}

A non-magnetic component is included to account for stray light from the instrumental system and unresolved mixing of adjacent magnetic and non-magnetic regions.  It is assumed the non-magnetic fill fraction is $1-f$.  A stray light profile is generated from averaged quiet sun intensity profiles and is allowed to shift in wavelength ($\lambda_s$), adding one additional free parameter.
\begin{figure*}
\centerline{\includegraphics[scale=0.69]{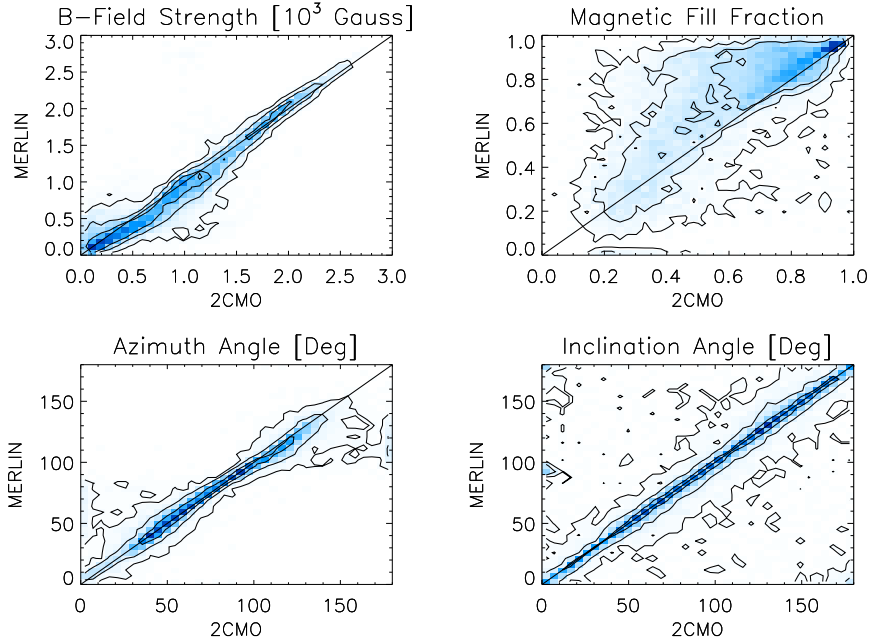}}
\caption{Scatter plot of the magnetic field parameters fitted by MERLIN compared with those from 2CMO from inversion of Hinode SOT/SP observations of NOAA 11024 taken on 2009-07-07.  Contours are drawn at 52, 84, and 97\% levels.  1:1 correspondence is indicated by the diagonal black line in each plot.}
\label{fig5}
\end{figure*}

To confirm the validity of the solutions from our inversion code, we have conducted a comparison of the results from 2CMO and those from the Milne-Eddington gRid Linear Inversion Network (MERLIN) from the Community Spectropolarimetric Analysis Center (CSAC) at the National Center for Atmospheric Research.  MERLIN is a new implementation of the original ME code written by \citet{skumanich87} \citep{lites07}.  MERLIN fits the same set of parameters as our code, and although the subtleties of the implementation, initialization, and calculation are different, the two codes describe the same physics and should produce similar results when performed on identical data.

As an example we have obtained MERLIN results from the CSAC inversion client for a 6302 \AA\ observation of the sunspot in NOAA 11024 taken on 2009 July 7 by the Solar Optical Telescope Spectro-Polarimeter onboard the Hinode spacecraft, and preformed a 2CMO inversion on the same dataset.  In Figure \ref{fig4} we show the observed Stokes profiles from a single position in the sunspot umbra (black), the fitted profiles from 2CMO (red) and MERLIN (blue), and the data-fit residuals.  The synthetic profiles show slight differences due to the choice of line parameters, however the quality of fitting shown by the data-fit residuals is comparable.  In Figure \ref{fig5} we show the scatter-plot comparison of the magnetic field parameters from MERLIN and 2CMO for all inverted positions.  The two codes produce largely the same magnetic field strength in the umbra to 70 G (rms).  The increased scatter in magnetic field strength at low values demonstrates the degeneracy with the magnetic filling factor in regimes where the line is not completely split.  However, the scatter in the filling factor is larger than we might expect based on this degeneracy alone, and in general larger values of the filling factor are produced by MERLIN.  While the parameter sets are identical between the two codes, the implementation is different, in particular the treatment of scattered light and the tolerances placed on the Stokes I fit can significantly alter the resulting fill factor.  The 2CMO inversion code and the results from the comparison are described further in \citet{jaeggliphdt}.

\subsection{Treatment of Stray Light}
\label{sec_stray_light}
Stray light is one of the largest sources of error in temperature and magnetic field measurements in sunspots.  Detailed discussions of the effects of stray light and its correction have been carried out in \citet{martinez93} and \citet{solanki93}.  In general, stray light results in reduced contrast of the continuum with respect to the quiet sun, the contamination of sunspot intensity profiles with a stray quiet sun component, and reduced amplitudes of the polarized components.  The correction of stray light is particularly important for the accurate determination of quantities which depend on intensity, such as temperature and the equivalent width of spectral lines.
\begin{figure}[!ht]
\centerline{\includegraphics[scale=0.5]{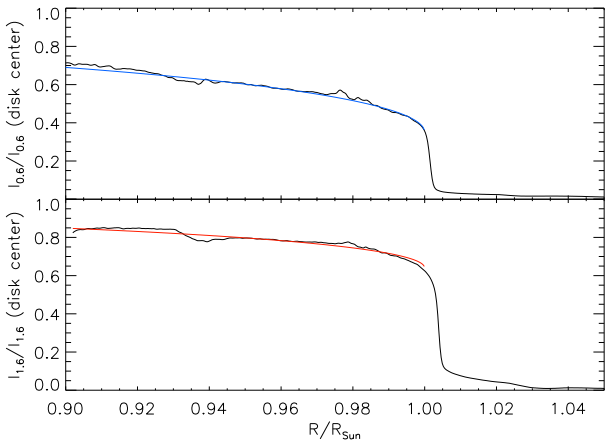}}
\caption{Profiles of the 6302 \AA\ (top) and 15650 \AA\ (bottom) continuum intensity at the solar limb as a function of distance from solar center.  The colored line in each case is the theoretical limb darkening function for 0.6 and 1.6 $\mu$m respectively.}
\label{fig6}
\end{figure}

Spatial stray light may arise from both atmospheric and instrumental scattered light.  Without proper characterization of the instrumental and atmospheric contribution to spatial stray light, the fraction due to the instrument and telescope cannot be distinguished from the unresolved mixing of adjacent magnetic and non-magnetic regions.  The degeneracy of magnetic field strength and its filling fraction in weak field regimes is an additional source of ambiguity for the estimation of the stray light somponent.  Therefore the filling fraction determined by the inversion can only place an upper limit on the spatial stray light.  Detailed measurements to determine the stray light for FIRS have not been carried out.  However, in observations at the solar limb (shown in Figure \ref{fig6}) the stray fraction at 0.02 R$_\odot$ or about 20" above the limb is approximately 3\% and 5\% of the disk center brightness in the visible and infrared (or about 8\% of the intensity at the limb).  This measurement agrees with the filling fraction determined in the umbra of very large sunspots, where the photosphere should be almost entirely magnetic.  Therefore we apply the inverted magnetic filling fraction to correct the spatial stray light in the intensity spectrum.

There is no line-free source of spatial stray light, therefore any spectral stray light must arise only following the slit of the instrument.  It is possible to estimate the spectral stray light by measuring the equivalent width of spectral lines and comparing them to a pristine measurement free of spectral stray light.  The presence of a possible large spectral stray fraction in FIRS and its effect on the measurement of equivalent widths is discussed further in \S\ref{sec_OH}.

\section{Inference of the \hh\ Abundance}
\label{sec_RT_model}
The direct measurement of \hh\ at photospheric heights is infeasible for the reasons described in \S\ref{sec_meth}.  Therefore, it is necessary to establish the \hh\ abundance through indirect methods.   The strength of the OH lines near 15650 \AA\ is strongly related to the temperature at the height of continuum formation, although the lines are formed higher in the atmosphere by about 100 km.  At high temperatures the behavior of the lines is adequately described by a simple collisional and radiative model \citep{penn03a}.  Therefore, they provide us with a way to characterize the state of the atmospheric gas in addition to the measurement of continuum intensity.  We establish the equivalent width of OH as a diagnostic for the \hh\ fraction using spectra synthesized from model atmospheres.

\subsection{Model Atmospheres}
In order to characterize the average properties of sunspot atmospheres we have examined 1-D atmospheric models with solar gravity and abundances, but with a range of effective temperatures.  We use the Kurucz model atmosphere grids (T$_{eff}$=4000 to 7000 K in 250 K steps) (http://kurucz.harvard.edu/grids.html), and to fully explore the full range of sunspot temperatures, cooler models from the updated Phoenix model atmosphere grids (T$_{eff}$=2600 to 3900 K in 100 K steps) were kindly provided by Peter Hauschildt \citep{hauschildt97}.
\begin{figure}
\centerline{\includegraphics[scale=0.5]{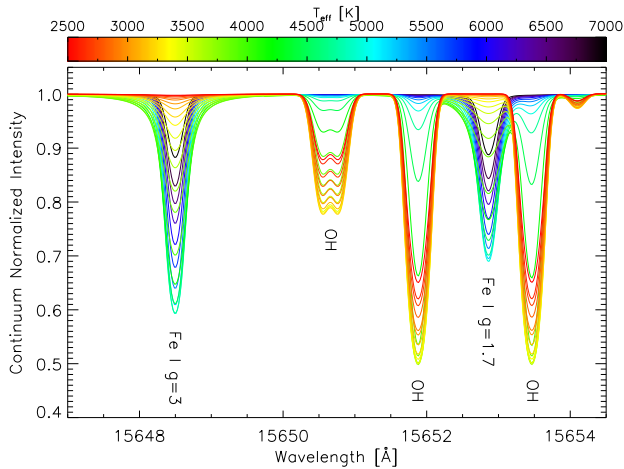}}
\caption{Synthetic spectra from the RH code for the FIRS 15650 \AA\ bandpass for each model atmosphere from an effective temperature of 2600 K (red) to 7000 K (violet).}
\label{fig7}
\end{figure}
\begin{figure}
\centerline{\includegraphics[scale=0.5]{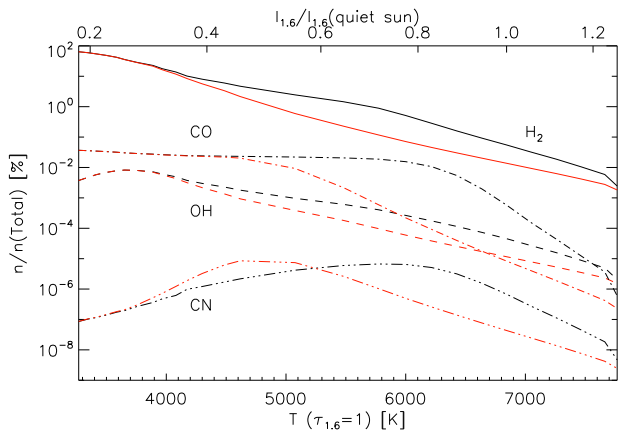}}
\caption{Semi-log plot of fraction of the total particle density made up by the \hh, CO, OH, and CN molecules at $\tau=1$ in the \ion{Fe}{1} 15648.5 \AA\ line core (black) and continuum (red) as a function of the 15650 \AA\ continuum temperature and intensity.}
\label{fig8}
\end{figure}

To generate the necessary spectral diagnostics, these model atmospheres were supplied to the Rybicky-Hummer (RH) code developed by Han Uitenbroek, which has been demonstrated in \citet{uitenbroek00a, uitenbroek00b}.  The code is able to solve the equations of radiative transfer and statistical and chemical equilibrium for atomic and molecular species in LTE and non-LTE situations.  We have used the LTE solutions from the code to generate synthetic spectra and obtain the necessary diagnostics for our 1-D plane parallel model geometry.  In addition to calculating the molecular populations at every height in the model, we used the RH code to produce a synthetic spectrum of the 15650 \AA\ range including the \ion{Fe}{1} and OH lines for each atmospheric model viewed at ten heliocentric angles (0.0$^\circ$, 10.2$^\circ$, 23.4$^\circ$, 36.2$^\circ$, 48.5$^\circ$, 60.0$^\circ$, 70.3$^\circ$, 78.9$^\circ$, 85.3$^\circ$, 89.1$^\circ$).   In Figure \ref{fig7} we show the continuum normalized spectra from disk center for each atmospheric model from an effective temperature of 2,600 K (red) to 7,000 K (violet).

The equivalent width of OH and the continuum intensity (normalized to the quiet sun model $T_{eff}$=5,750 K) were measured directly from the synthetic spectrum for each model and heliocentric angle.  The corresponding molecular abundances were determined at two heights in the model atmosphere, $\tau=1$ in the continuum where the intensity/temperature is measured, and at $\tau=1$ in the \ion{Fe}{1} line core where the magnetic field strength measurement can be considered to originate \citep{sanchez96}.  A linear interpolation was used to retrieve the parameters for arbitrary heliocentric angles and temperatures.
\begin{figure}
\centerline{\includegraphics[scale=0.5]{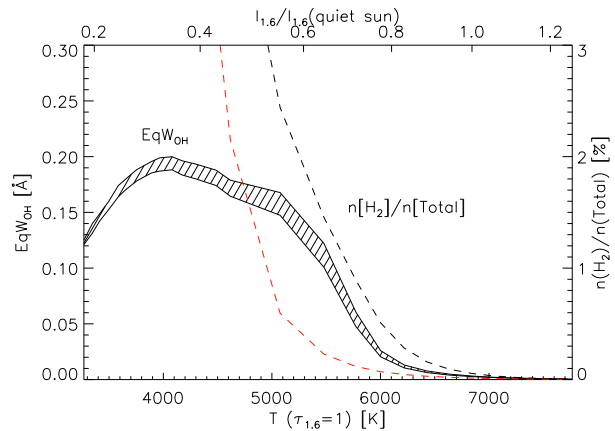}}
\caption{The equivalent width of the synthesized OH 15651.9 \AA\ line plotted against the 15650 \AA\ continuum temperature and intensity.  The \hh\ density fraction from Figure \ref{fig8} on a linear scale is included for reference and should be read from the right hand scale.}
\label{fig9}
\end{figure}
\begin{figure}
\centerline{\includegraphics[scale=0.5]{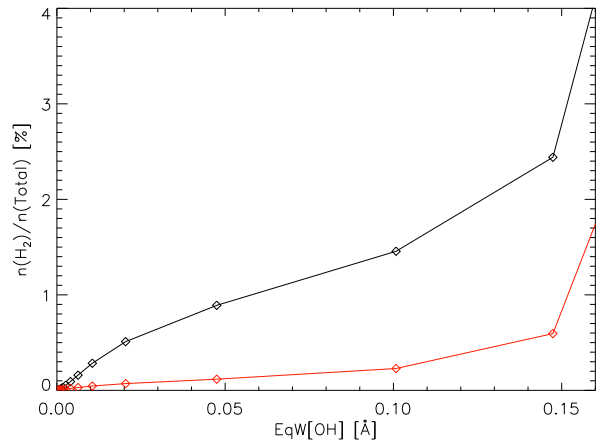}}
\caption{The \hh\ fraction at the $\tau=1$ height for the the \ion{Fe}{1} 15650 line core (black) and continuum (red) as a function of OH equivalent width.}
\label{fig10}
\end{figure}

Figure \ref{fig8} shows the logarithmic fraction of the total particle number of the molecules \hh, OH, CO, and CN at $\tau=1$ in the 15648.5 \AA\ \ion{Fe}{1} line core (black lines) and continuum (red lines) as a function of the continuum temperature and intensity at disk center. The OH fraction closely tracks the increase of the \hh\ fraction from about 6,500 to 3,500 K. Below 3,500 K the OH abundance decreases due to competition with CO for atomic oxygen. The other molecular species do not follow the behavior of \hh\ particularly well. This demonstrates the validity of OH as a proxy for \hh\ for atmospheres with temperatures above 3,500 K.  Figure \ref{fig9} shows the predicted equivalent widths of the OH 15651.9 \AA\ line over the range of heliocentric angles spanned by the sample as a function of the 15650 \AA\ continuum temperature and intensity at disk center (black shaded region).  The fractional number density of \hh\ for the line core (red dashed line) and continuum (black dashed line) are included in linear scale to show the good proportionality between OH equivalent width and \hh\ density.  Figure \ref{fig10} shows the direct relation between OH equivalent width and the \hh\ fraction at disk center.

It is necessary to use multiple heliocentric angles to determine the correct relation between the equivalent width of the 15651.9 \AA\ OH line and \hh.  Because the height of $\tau=1$ changes as a function of heliocentric angle, the OH line strength, intensity ratios, and atmospheric parameters for sunspots at significantly different positions on the solar disk cannot be compared directly.  To make use of observations at obtained over a large range of heliocentric angles we must apply a correction based on the atmospheric models.  For the heliocentric angle of each sunspot observation we determine the relation of the intensity and OH equivalent width at the heliocentric angle to the equivalent disk center continuum temperature and \hh\ fraction respectively.  In this way data from a wide range of angles can be compared directly.

\subsection{Measurement of EqW$_{\rm OH}$}
\label{sec_OH}
We have chosen to use the 15651.9 \AA\ OH line which is stronger than the blend at 15651.0 \AA, and less affected by the Zeeman split \ion{Fe}{1} g=1.7 line than the 15653.4 \AA\ line.  However, the presence of the g=1.7 line still significantly affects the inferred equivalent width of the OH line at even modest magnetic field strengths.  Observational systematics in FIRS such as fringing and improper correction of the filter profile increase the uncertainty in the direct measurement of equivalent width from the integrated spectrum.  Therefore, in the umbra we calculate the equivalent width from the inverted parameters for the OH line, and elsewhere in the sunspot the equivalent width is calculated directly from the Stokes I spectrum because it does not produce poor results when the lines are weak.

\begin{figure}
\centerline{\includegraphics[scale=0.45]{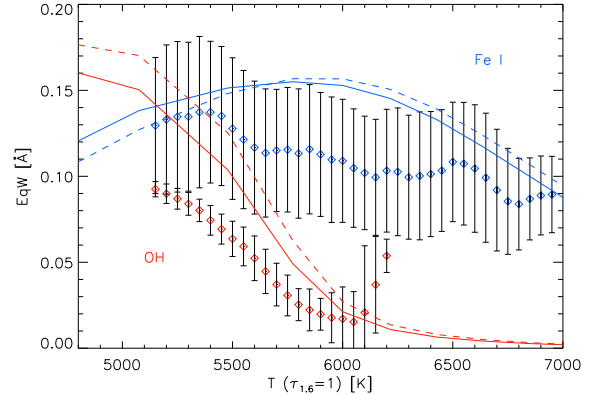}}
\caption{Equivalent width of the OH 15651.9 \AA\ (red) and \ion{Fe}{1} 15654.8 g=3 (blue) lines compiled from the  sample vs. continuum temperature at 15650 \AA\ are indicated by the points with error bars.  The theoretical values of the OH and \ion{Fe}{1} line equivalent widths for heliocentric angles spanning the sample are indicated by the solid (disk center) and dashed lines ($\sim50^\circ$).}
\label{fig11}
\end{figure}

Figure \ref{fig11} shows the compiled EqW$_{OH}$ vs. continuum temperature for the sample.  The data have been arranged into 50 K bins and the mean and standard deviation of each bin was calculated.  The data and model describe qualitatively the same curve which appears to be different only by a factor of about 1.55.

This discrepancy between the empirical and theoretical relation of OH equivalent width with temperature must come from systematics in the observations or a lack of realism in the model atmospheres and the RH code.  A large amount of spectral stray light could account for the consistent underestimation of the OH equivalent width.  If we assume a model where the stray light level at any given position is constant fraction of the quiet sun continuum ($I'=I+f*I_{0}$, e.g. from uniform scattered or parasitic light), then the corrected equivalent width would be increased, but the continuum contrast would also be increased, and the resulting relation would move up and to the left in Figure \ref{fig11}.  However, if we assume that the stray fraction at any given position is proportional to the continuum intensity at that position ($I'=I*(1+f)$, e.g. from overlapping orders), then the equivalent width increases by a multiple of the stray fraction, but the continuum contrast is preserved.

If there were a large fraction of spectral stray light it should affect the iron lines in the same way.  The equivalent width of the whole Zeeman triplet was calculated for the sunspots and the surrounding regions of quiet sun.  The data was binned and the mean and standard deviation were determined in the same way as for the OH line, and is indicated by the blue points in Figure \ref{fig11}.  The large scatter in the \ion{Fe}{1} measurement makes the possible stray light contamination difficult to determine for these lines, but it appears that they also underestimate the \ion{Fe}{1} equivalent width relative to the model.

We are reasonably confident with the calculation of the OH populations and the lines in the synthetic spectra, however the same cannot be said about the \ion{Fe}{1} lines.  The \ion{Fe}{1} lines have non-LTE ionization and we have calculated them assuming LTE.  The infrared NSO sunspot atlas (atlas 5) obtained with the Fourier Transform Spectrometer (FTS) on the McMath Solar Telescope at Kitt Peak provides a high-resolution, pristine solar spectrum with exceptionally low spectral stray light \citep{wallace01}.  From the profiles (available at ftp://nsokp.nso.edu/pub/atlas/) we have measured an equivalent width of 0.0968 \AA\ for the \ion{Fe}{1} 15648.5 \AA\ g=3 line which is in very good agreement with the model at the quiet sun continuum temperature of 6850 K.  In spite of this agreement, we can trust neither the measurement of \ion{Fe}{1} equivalent width, nor the RH model for the same quantity beyond the quiet sun, and we must rely on OH.
\begin{figure}
\centerline{\includegraphics[scale=0.45]{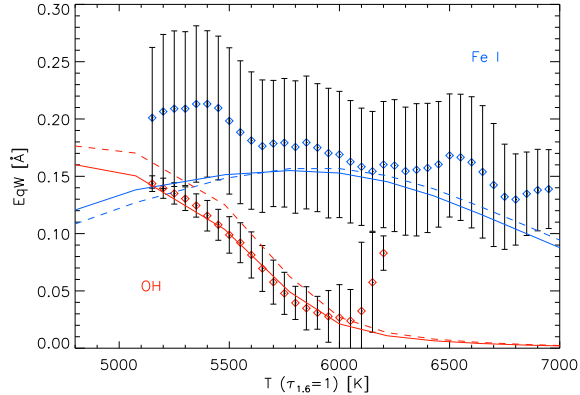}}
\caption{The same as Figure \ref{fig11}, but the equivalent widths measured from the sample have been scaled by a factor of 1.55.}
\label{fig12}
\end{figure}

The OH and \ion{Fe}{1} equivalent widths, adjusted by a factor of 1.55 in comparison to the models, are shown in Figure \ref{fig12}.  If we are to believe this factor, it implies a spectral stray light fraction of 35\%.  This model of stray light is not particularly realistic for FIRS, where we might expect overlapping orders to fall onto neighboring slits.  From the favorable comparison of FIRS 6302 \AA\ with Hinode SOT/SP data, which shows very comparable contrast in both the continuum and in the lines, we can conclude that the possible stray fraction is at least limited to infrared.  In the future, further verification of the RH synthesis of OH is possible using OH lines at visible wavelengths.  Additional investigation is necessary to confirm the presence of such a large spectral stray light fraction in FIRS, but for the moment we apply a universal factor of 1.55 to the OH equivalent widths and proceed with the \hh\ inference under this caveat.

\section{Selected Survey Results}
\label{sec_results}
In this section we provide context for the observations using sunspot histories that have been determined from SOHO/MDI and SDO/HMI daily intensitygrams, and discuss results for each sunspot individually.  The sunspot observations have been inverted to retrieve the magnetic field, the OH equivalent has been measured, and model atmospheres have been applied to infer the continuum temperature and \hh\ fraction.

\begin{figure*}[!ht]
\centerline{\includegraphics[scale=0.7]{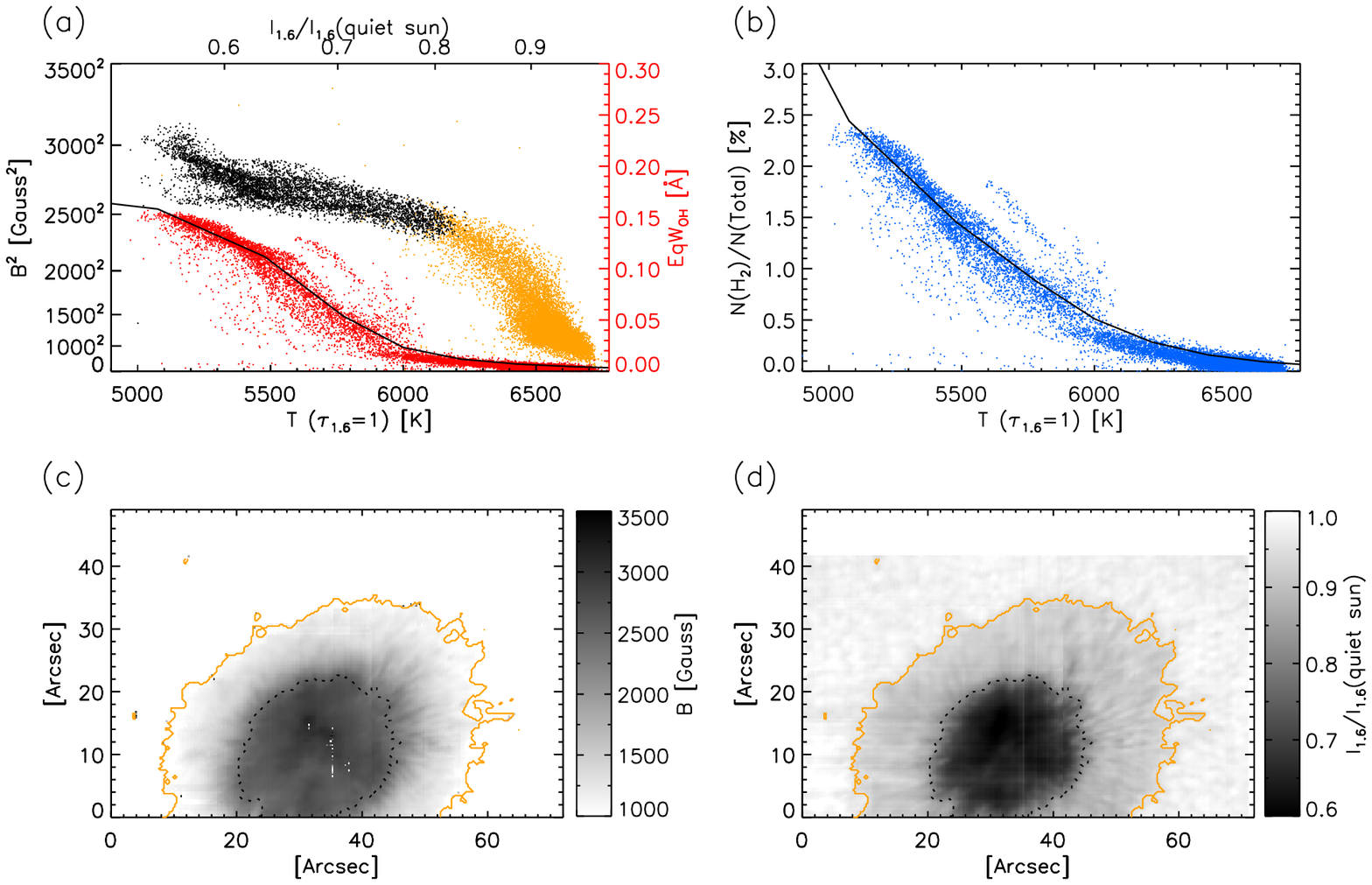}}
\caption{For NOAA 11131 taken on 2010-12-06 (a) thermal-magnetic relation and OH equivalent width, (b) \hh\ fraction at the height of the \ion{Fe}{1} line core inferred from the OH equivalent width vs. the continuum temperature, (c) magnetic field map, (d) map of continuum intensity.}
\label{spot1}
\end{figure*}

\begin{figure*}[!ht]
\centerline{\includegraphics[scale=0.7]{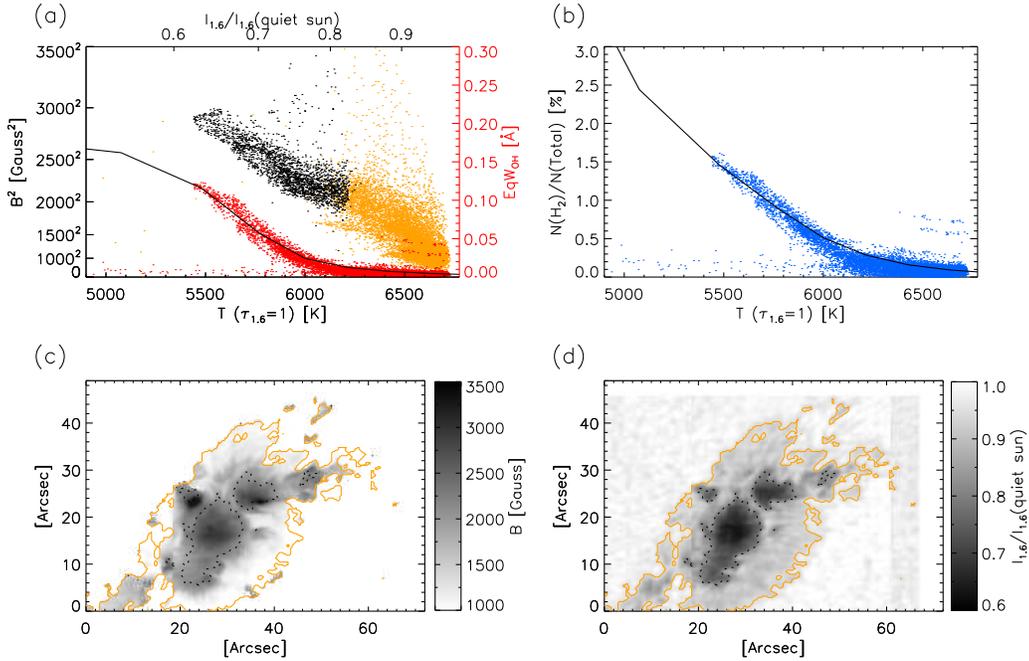}}
\caption{For the leading polarity spot in NOAA 11035 taken on 2009-12-17 (a) thermal-magnetic relation and OH equivalent width, (b) \hh\ fraction at the height of the \ion{Fe}{1} line core inferred from the OH equivalent width vs. the continuum temperature, (c) magnetic field map, (d) map of continuum intensity.}
\label{spot2}
\end{figure*}

\begin{figure*}[!ht]
\centerline{\includegraphics[scale=0.7]{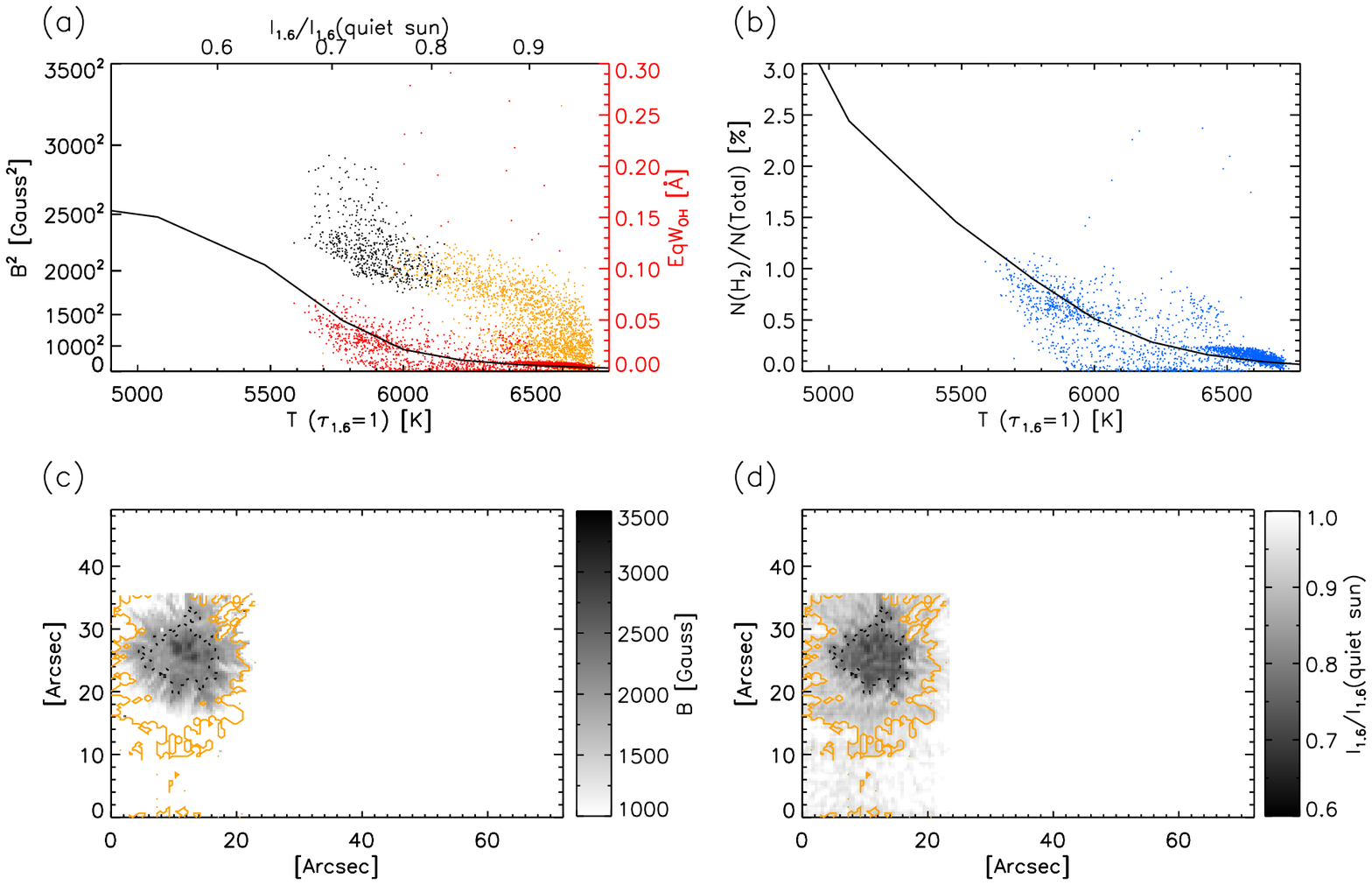}}
\caption{For NOAA 9429 taken on 2001-04-18 (a) thermal-magnetic relation and OH equivalent width, (b) \hh\ fraction at the height of the \ion{Fe}{1} line core inferred from the OH equivalent width vs. the continuum temperature, (c) magnetic field map, (d) map of continuum intensity.}
\label{spot4}
\end{figure*}

\begin{figure*}[!ht]
\centerline{\includegraphics[scale=0.7]{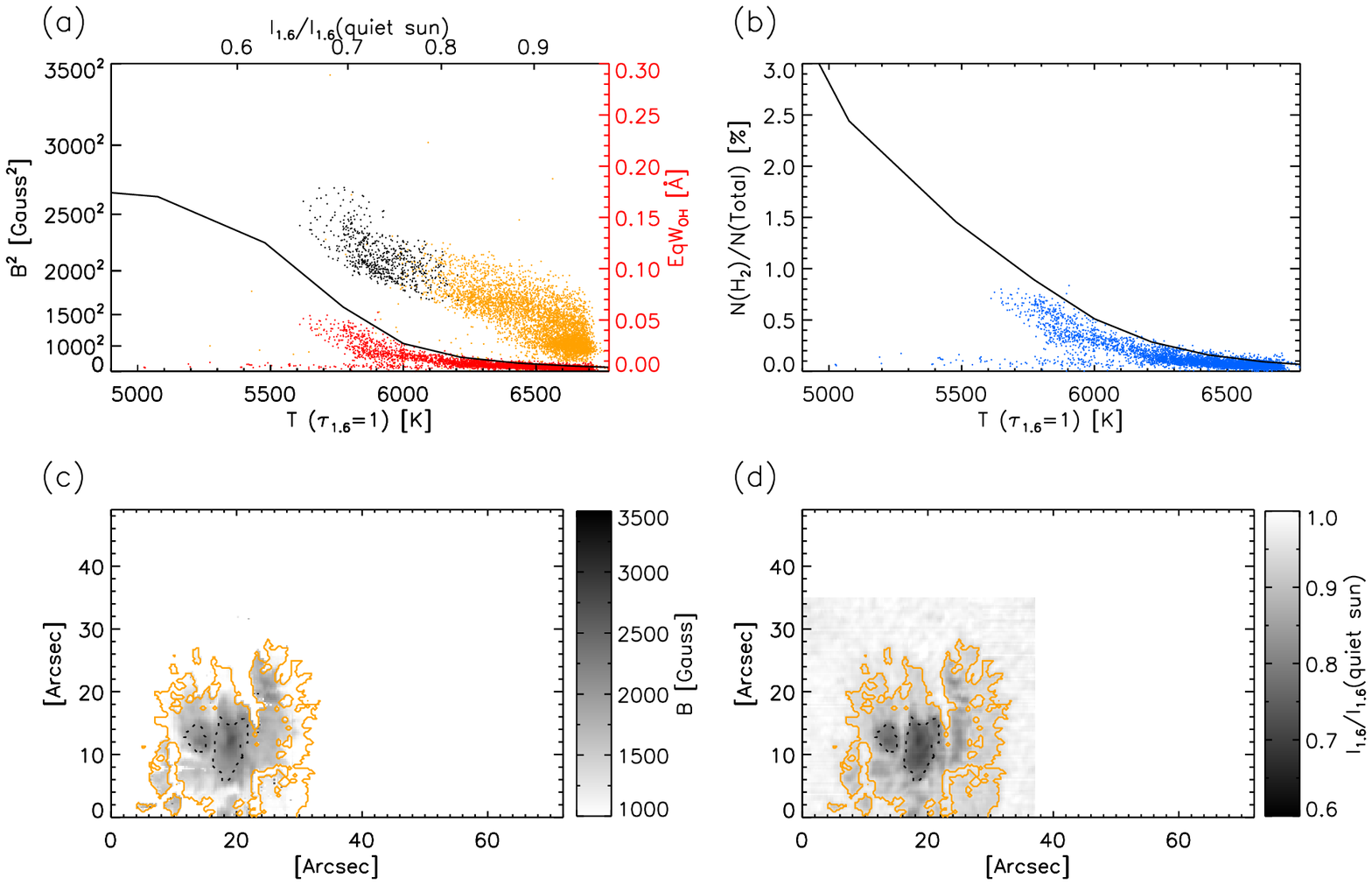}}
\caption{For NOAA 11130 taken on 2010-12-02 (a) thermal-magnetic relation and OH equivalent width, (b) \hh\ fraction at the height of the \ion{Fe}{1} line core inferred from the OH equivalent width vs. the continuum temperature, (c) magnetic field map, (d) map of continuum intensity.}
\label{spot5}
\end{figure*}

\begin{figure*}[!ht]
\centerline{\includegraphics[scale=0.7]{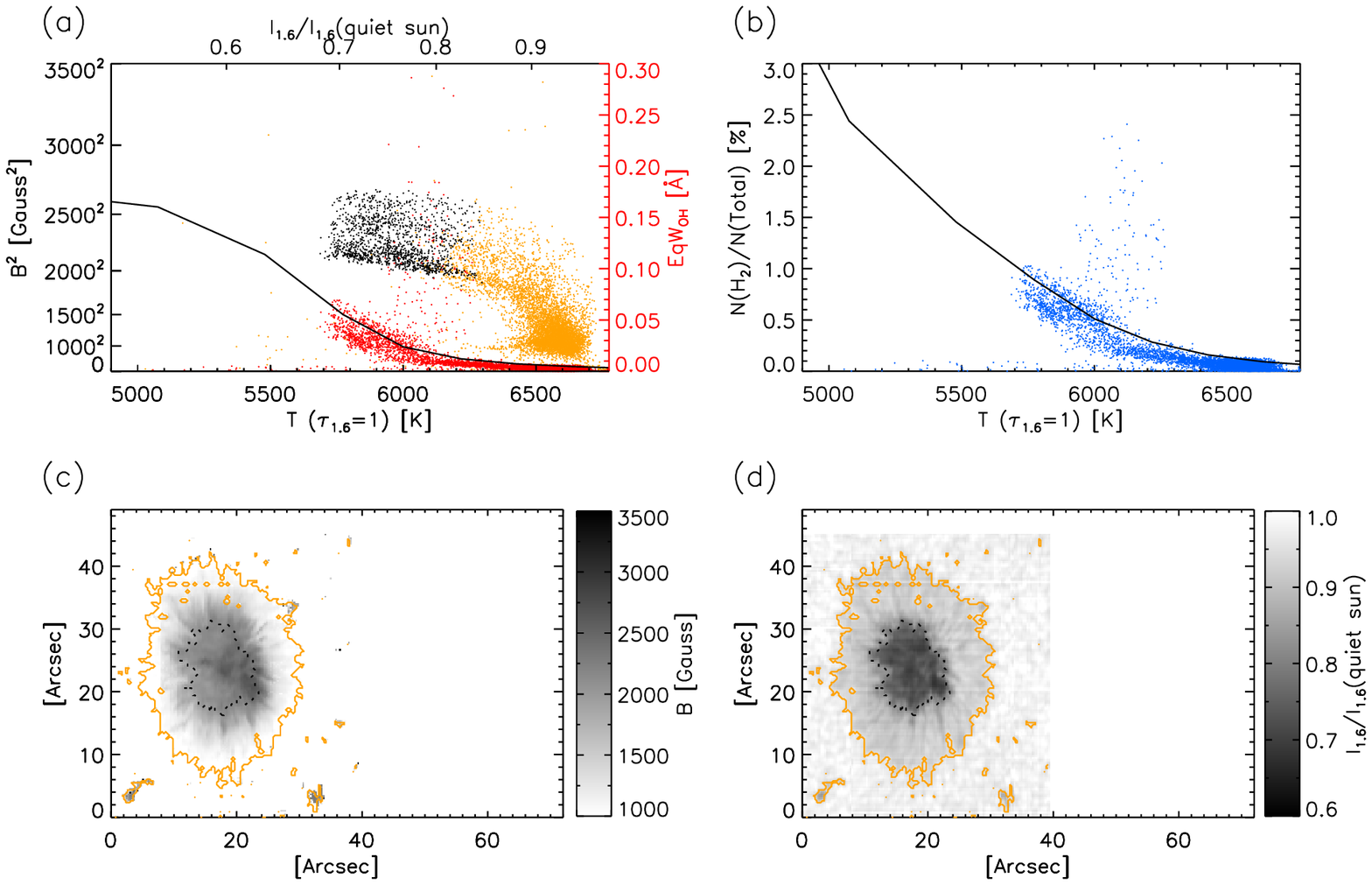}}
\caption{For NOAA 11101 taken on 2010-09-02 (a) thermal-magnetic relation and OH equivalent width, (b) \hh\ fraction at the height of the \ion{Fe}{1} line core inferred from the OH equivalent width vs. the continuum temperature, (c) magnetic field map, (d) map of continuum intensity.}
\label{spot3}
\end{figure*}

\begin{figure*}[!ht]
\centerline{\includegraphics[scale=0.7]{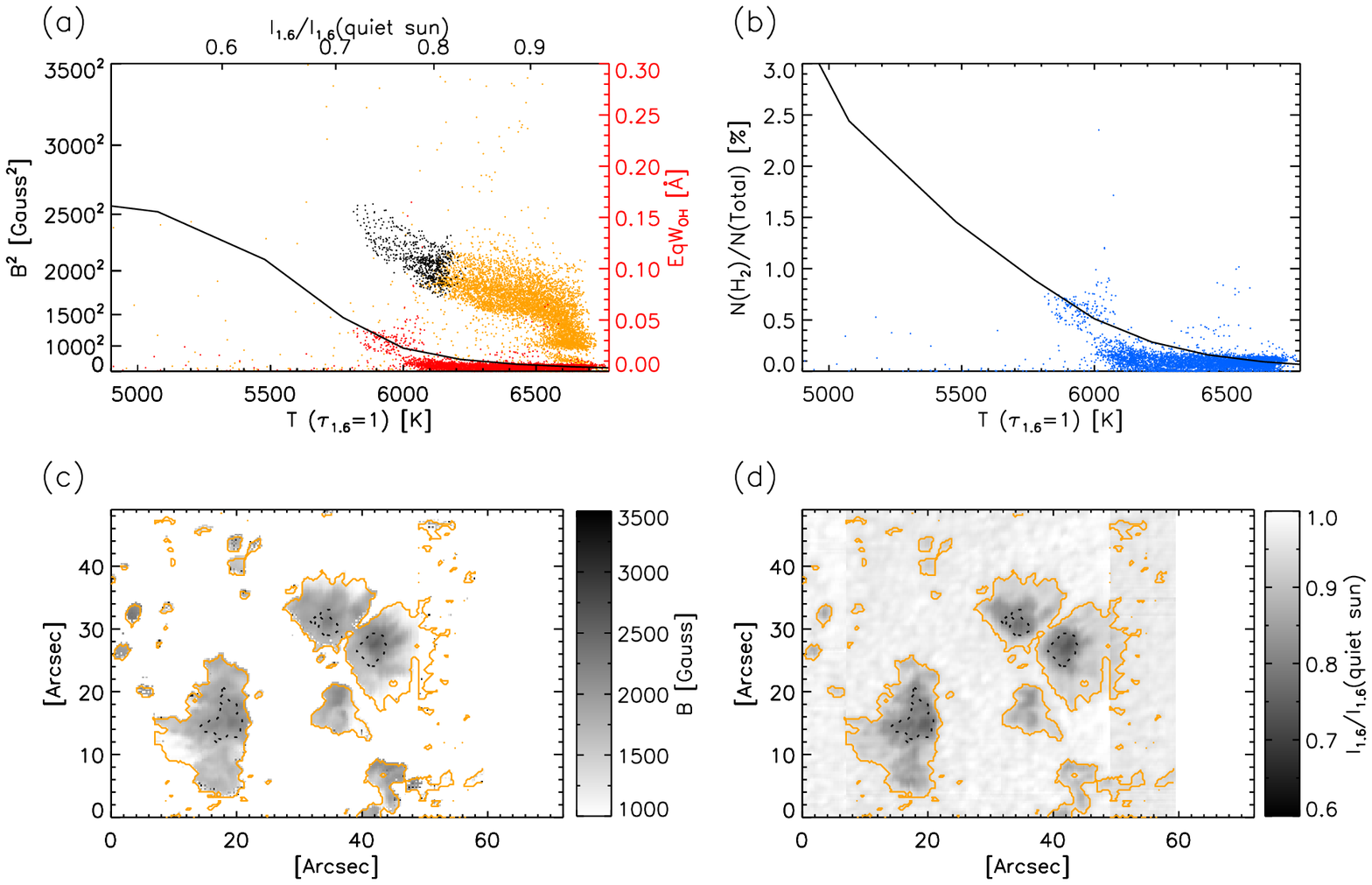}}
\caption{For NOAA 11046 taken on 2010-02-13 (a) thermal-magnetic relation and OH equivalent width, (b) \hh\ fraction at the height of the \ion{Fe}{1} line core inferred from the OH equivalent width vs. the continuum temperature, (c) magnetic field map, (d) map of continuum intensity.}
\label{spot6}
\end{figure*}

\begin{figure*}[!ht]
\centerline{\includegraphics[scale=0.7]{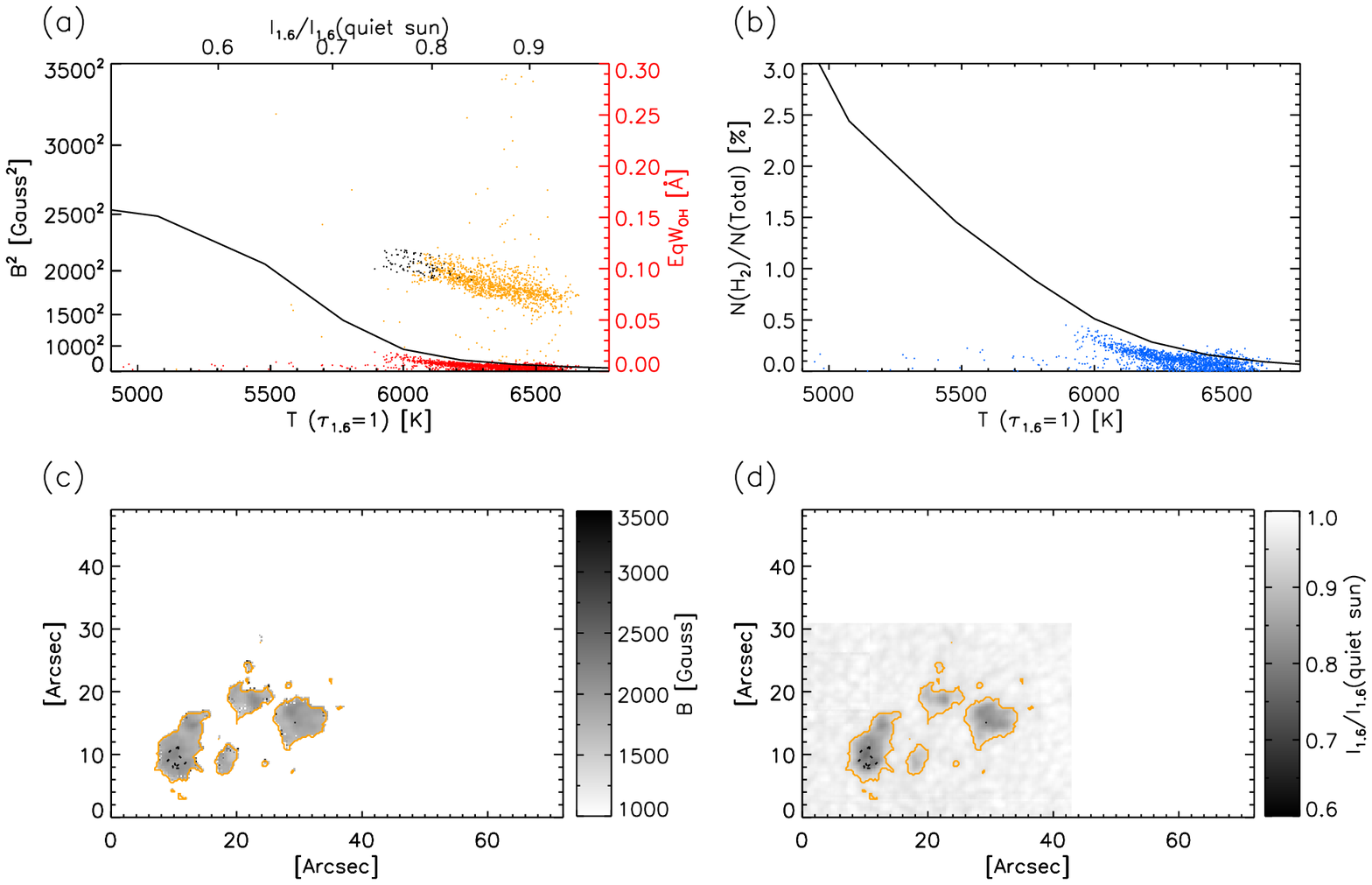}}
\caption{For NOAA 11049 taken on 2010-02-19 (a) thermal-magnetic relation and OH equivalent width, (b) \hh\ fraction at the height of the \ion{Fe}{1} line core inferred from the OH equivalent width vs. the continuum temperature, (c) magnetic field map, (d) map of continuum intensity.}
\label{spot7}
\end{figure*}

The results from this analysis for each of our example sunspots are shown in Figures \ref{spot1} to \ref{spot7} sorted from strongest to weakest by umbral magnetic field strength.  In panel (a) for each spot observation we show the thermal-magnetic relation (square of the magnetic field plotted against the continuum brightness temperature), umbral (black points) and penumbral (gold points) regions have been selected based on intensity.  In the same panel we show the measured OH equivalent width scaled by a factor of 1.55 and the OH equivalent width for the sunspot heliocentric angle produced by the RH code (black line).   In panel (b) of the figure we show the inferred \hh\ fraction at the height of $\tau=1$ in the \ion{Fe}{1} 15648.5 \AA\ line core for a disk center model atmosphere plotted against the continuum temperature in blue, the theoretical relation between temperature the \hh\ fraction is included in black.  Maps of the magnetic field strength and continuum intensity are shown in panels (c) and (d) respectively.  The solid gold line and dotted black line in each of the maps show the selected penumbral and umbral boundaries respectively.

\begin{deluxetable}{lccc}
\tablecaption{Summary of Selected Results \label{tbl2}}
\tablewidth{0pt}
\tablehead{
\colhead{Target Name} & \colhead{Max. B [G]} & \colhead{Min. T [K]} & \colhead{Max. N$_{H_2}$/N$_{Total}$ [\%]}
}
\startdata
NOAA 11131	& 3100 & 5100 & 2.3 \\ 
NOAA 11035	& 3000 & 5450 & 1.5 \\ 
NOAA 9429	& 2900 & 5700 & 1.0 \\ 
NOAA 11130	& 2700 & 5650 & 0.8 \\ 
NOAA 11101	& 2600 & 5750 & 0.9 \\ 
NOAA 11046	& 2500 & 5850 & 0.7 \\ 
NOAA 11049	& 2200 & 5950 & 0.3 \\ 
\enddata
\end{deluxetable}


\begin{description}
\item[NOAA 11131] was a large $\alpha$-spot that appeared fully formed from around the east limb on 1 Dec 2010, remained roughly the same size and shape until it disappeared around the west limb on 13 Dec.  The region developed a few transient pores during its passage of the disk. The observation in Figure \ref{spot1} was taken two days before it crossed the central meridian.

NOAA 11131 is the largest spot in the sample with a maximum umbral field of 3,100 G.  The thermal-magnetic relation follows a significantly different path from the other spots.  Large magnetic field strengths are reached early in the penumbra followed by a shallow slope through a majority of the umbra.  An abrupt increase in slope occurs at a temperature of 5,300 K.  The OH lines are quite strong, reaching a value corresponding to 2.3\% of hydrogen bound into \hh.

\item[NOAA 11035] was a complex $\beta\gamma$-region that emerged rapidly on 14 Dec 2009, $10^\circ$ from the central meridian.  Disorganized pore groups coalesced into large leading and following polarity sunspots.  The observation of the leading spot in Figure \ref{spot2} was taken on 17 Dec when the spots were largest.  After producing several C-class flares on this day the region began to decay, but the spots remained large as they traveled around the west limb on 22 Dec.

The NOAA 11035 observation shows a maximum umbral magnetic field strength of 3,000 G in the main umbra.  The spot shows exceptionally high magnetic field strengths in excess of 3,500 G in small features located to the upper left and upper right of the main umbra (can be clearly identified in panel (c)).  Interestingly these strong magnetic fields appear at the umbra/penumbra boundary of the secondary umbrae and do not correspond with the coolest temperatures, or strongest OH lines, and are probably a manifestation of flare activity.  Disregarding the scatter in magnetic field caused by these discrete magnetic features, the main thermal-magnetic relation for this spot does not display a sharp upturn, the relation gradually increases in slope in the main umbra starting at 5,900 K.  The \hh\ fraction inferred from the OH equivalent width is in good agreement with the model and reaches a maximum value corresponding to a \hh\ fraction of about 1.5\%.

\item[NOAA 9429] was an alpha spot that appeared fully formed around the east limb on 13 Apr 2001 and disappeared around the west limb on 25 Apr.  The observation in Figure \ref{spot4} was taken on 18 Apr just before it passed disk center, following this it appeared to slowly decay.

This is an early observation taken with the HSG prior to the commissioning of FIRS.  The NICMOS 3 detector used for this observation appears to have a significantly non-linear response which was not measured at the time of observation.  However, the umbral magnetic field strength derived from Stokes profile inversion is independent from intensity and still a reliable quantity.  We have adjusted the intensity linearly to match the FIRS observations, and the intensity-dependent quantities, (temperature, OH equivalent with, and N(\hh)/N(Total)) should only be considered in a qualitative way.  Despite the uncertainty in the temperature scaling, this sunspot shows a striking isothermal increase in magnetic field up to strengths of 2,900 G at the lowest temperatures.  This upturn can be ascribed to three strong magnetic cores visible in the center of the sunspot umbra seen in panel (c) of Figure \ref{spot4}.

\item[NOAA 11130] was a rapidly evolving, medium sized $\beta$-region with a single lead spot and weaker following group which emerged near disk center on 27 Nov 2010.  The region reached a maximum size on the 30th, the lead spot having developed a full penumbra, and following spots with partial penumbrae.  It slowly decayed over the next 4 days as it reached the limb.  The observation shown in Figure \ref{spot5} was taken of the leading polarity sunspot on 2 Dec during this decay phase when the spot group with following polarity had diminished to pores.

While NOAA 11130 does not have a particularly strong magnetic field (2,600 G) or a large \hh\ fraction, this sunspot also shows a clear case of isothermal intensification of the umbral magnetic field.  The magnetic field turns sharply upward, increasing by approximately 500 G starting at 5,800 K.  The inferred \hh\ fraction for this sunspot in particular is in rather poor agreement with the theoretical prediction, with a maximum of 0.8\% \hh.

\item[NOAA 11101] appeared around the east solar limb as a fully formed $\alpha$-spot with a penumbra on 25 Aug 2010 and did not experience significant evolution as it transited the disk and disappeared around the west limb on 5 Sep.  The observation shown in Figure \ref{spot3} was taken just after it had passed disk center on 2 Sep.

This sunspot shows a large scatter in the thermal-magnetic relation from an extended magnetic structure which spans the umbra-penumbra boundary on the lower right hand side of the sunspot umbra.  However, it shows a higher density of points at 5,800 K rising isothermally out of the main B$^2$-T track.  Agreement with the theoretical \hh-temperature relation is excellent and we can infer a maximum \hh\ fraction of 0.9\%.

\item[NOAA 11046] appeared around the east limb as a small $\beta$-region on 7 Feb 2010.  It became more developed around 11 or 12 Feb with a larger following polarity group and a smaller leading polarity sunspot.  The observation in Figure \ref{spot6} is of the following spot group on the 13th as the region began to decay over the next 5 days.  The spots seemed to have disappeared almost entirely as it reached the limb on 18 Feb.

Although this sunspot group is composed of three umbrae with partial penumbrae, it displays a cohesive thermal-magnetic relation which shows a clear increase in slope in the umbra, at about 6,100 K.  A maximum umbral field strength of 2,500 G is reached in the right-most umbra.  From the OH measurement we can infer a maximum \hh\ fraction of just under 0.7\%.

\item[NOAA 11049] was a small $\beta$-region that emerged on 16 Feb 2010 about $15^\circ$ east of the central meridian.  Although the region continued to evolve, the lead and following pores remained roughly equivalent in size until 21 Feb when the lead pore became stronger and the following region diminished before the region disappeared around the west limb on 24 Feb.  The observation shown in Figure \ref{spot7} of the following polarity pore group was taken on 19 Feb during this relatively quiescent evolution stage.

This pore group shows a very linear thermal-magnetic relation and low magnetic field strengths, a 2,200 G maximum is reached in the left-most pore.  The OH lines are small or undetected, corresponding to 0.3\% \hh\ fraction at most.

\end{description}

For the sample as a whole, the overall agreement of the scaled OH equivalent width with the theoretical OH curve, especially in the larger sunspots, strengthens our confidence in the results of the radiative transfer and chemical equilibrium code.  The maximum umbral field strength occurs at the same location as the maximum OH equivalent width (except for the flaring region, NOAA 11035) as we would expect if OH is strongly linked to atmospheric temperature, and temperature is controlled directly by magnetic suppression of the transport of convective energy.

Among the sample, the medium-sized spots NOAA 9429 and 11130 show clear isothermal increases in the magnetic field strength which occur around T=5,800 K.  NOAA 11101 shows a larger scatter of the points in the B$^2$ vs. T curve, with a higher density track similar to those of NOAA 9429 and 11130, it is suggestive of the existence of isothermal magnetic field intensification process.  For the larger spots, NOAA 11131 and 11035, and for the smaller fragmented sunspot group NOAA 11046, a clear case for isothermal magnetic field intensification cannot be made.  However, it should be noted that all of the larger sunspots show some kind of increase in the slope of the thermal-magnetic relation starting below 6000 K where the \hh\ fraction begins to increase rapidly.  The pores in NOAA 11049 show strictly linear behavior, indicating that for the case of small pores, the magnetic field is too weak and the temperature too high to have a significant \hh\ fraction, which is confirmed by the undetected OH lines.

A very intriguing result of this survey is the B$^2$ vs. T curve of the large spot of NOAA 11131 that is distinctly different from those of the rest of the sample. It follows an elevated track in B$^2$ vs. T without the signature of isothermal magnetic field intensification at T = 5,800 K. Similar results are found in the other large sunspots in the sample.

\section{Conclusions and Discussion}
\label{sec_conclusions}
Our survey and analysis provide observational evidence that significant \hh\ molecule formation is present in sunspots that are able to maintain maximum fields of greater than 2,500 G.  Measurements of the OH equivalent width seen in sunspot umbrae are qualitatively consistent with the predictions from spectra synthesized by radiative transfer models.  We infer a molecular gas fraction of a few percent \hh\ in the largest sunspots.  The formation of this small faction appears to alter the equilibrium of pressures in the sunspot isothermally, resulting in an increase of the slope of the thermal-magnetic relation at temperatures lower than a 15650 \AA\ continuum brightness temperature of 6,000 K where the \hh\ fraction begins to rapidly increase with temperature.  We suggest that the formation of \hh\ molecules in the sunspot umbra causes a rapid intensification of the magnetic field without a significant decrease in temperature which would explain the increase in slope of the thermal-magnetic relation.

We hypothesize that \hh\ plays an important role in the formation and evolution of sunspots.  During the initial stage of sunspot emergence and cooling, the formation of \hh\ may trigger a temporary ``runaway'' magnetic field intensification process.  As magnetic flux emerges and strengthens, the sunspot atmosphere cools due to the suppression of convective heating by the magnetic field.  When sufficiently low temperatures are reached \hh\ begins to form in substantial numbers in the coolest parts of the umbra.  As free hydrogen atoms combine to form \hh, the total particle number density is reduced.  The dissociation energy released into the atmosphere is rapidly dissipated by radiative cooling due to the low opacity of the photosphere, reducing the total kinetic pressure without a corresponding reduction in temperature.  The decrease in gas pressure causes this region to shrink in size, and due to the high electrical conductivity of the atmosphere the magnetic fields are compressed with the plasma (the ``frozen-in field'' effect). The resulting higher magnetic field density further inhibits the convective heating of the sunspot atmosphere, which leads to further cooling. This ``runaway'' magnetic field intensification process is most likely a temporary phenomenon which is arrested long before all the hydrogen atoms have condensed into molecular form.  At some point the transport of energy by convection will be effectively quenched and increases in magnetic field will no longer result in decreases in temperature.  The transfer of radiative energy from surrounding hotter regions would also keep the umbra from becoming excessively cool. Therefore further \hh\ formation would be halted.

While the formation of \hh\ may initiate a more rapid intensification of the sunspot magnetic field during sunspot emergence, we speculate that during the decay phase in a sunspot which has already formed a substantial \hh\ population, the highest concentrations of molecular gas would tend to maintain the magnetic field against decay and extend the lifetime of the sunspot.  As the magnetic field in a sunspot weakens, regions of the umbra once cool become warm and \hh\ dissociates back into atomic hydrogen.  The more rapid increase in pressure in warmer regions of the umbra would compress the remaining cool regions, concentrating the magnetic field and maintaining the cool interior against convective heating.

The formation of \hh\ would speed up the process of sunspot emergence, and the dissociation of \hh\ would slow down their disappearance.  While the effects of the formation and destruction of molecules would produce similar signatures in the thermal-magnetic relation, we would expect to see more cases of sunspots in the decay phase due to observational bias.  There is evidence that the intensification of the magnetic field occurs in discrete cores, such as can been seen in NOAA 9429, which is consistent with this speculative picture of molecule formation during the growth and decay of sunspots.  If this is the case the molecular fraction would be significantly underestimated in sunspots due to the effect of filling factor, and may exist in quantities of 5\% in unresolved features, consistent with the coolest models of the umbra such as those presented in \citet{maltby86}.

It is possible that two other effects contribute to the non-linearity of the thermal-magnetic relation.  In previous studies \citep{martinez93,solanki93,mathew04}, the non-linearity of the thermal-magnetic relation was interpreted as a radiative transfer effect, i.e. the Wilson Depression effect in sunspots.  Cooler atmospheres in a sunspot are more optically thin, therefore the magnetic field and continuum measurements originate from a greater geometrical depth.  In seeing deeper into the atmosphere we are able to see  relatively hotter regions, therefore the observed temperature would seem to decrease less rapidly (relative to radius or B) than in a single geometrical layer.  This effect would therefore tend to increase the slope of the thermal-magnetic relation, however the temperature should still decrease as the magnetic field strength increases.  

Through all of this work we have also neglected the curvature force.  Increased contributions to the horizontal support from the curvature force in outlying regions may cause the magnetic pressure in the sunspot core to seem boosted, contributing to the non-linearity of the thermal-magnetic relation.  

Detailed modeling efforts are necessary to determine the validity of the proposed scenarios for \hh\ formation and destruction during the emergence and decay of sunspots, and contribution of the radiative transfer effect and the curvature force to the non-linearity of the thermal-magnetic relation.  We intend to more carefully consider the effects of the curvature force and the Wilson Depression and further investigate the problem of \hh\ formation using the simultaneous dual-height observations obtained with the 6302 and 15650 \AA\ channels of FIRS to perform a detailed comparison with recent MHD sunspot models from \citet{rempel10}.

Particularly for the cases of isothermal intensification of the magnetic field in NOAA 9429 and 11130, it is unlikely that such a sharp upturn in the slope can be explained simply through radiative phenomenon or the curvature force.  The formation of \hh\ appears to be the most likely cause of the sharp increase in slope of the thermal-magnetic relation in sunspots.  In addition to this magnetic intensification process, the formation of molecules increases the heat capacity of the sunspot atmosphere.  Due to the additional degrees of freedom of the \hh\ molecule, the formation of a \hh\ fraction of 10\% would ideally raise the heat capacity of the gas by 13\% over an equivalent number density of atomic gas.  This non-thermal reservoir for energy may have an important effect on the local radiative output of the Sun.  Consequently, we suggest that modeling of the MHS equilibrium condition of sunspots in the form of Equation \ref{eqn1} must include a multiple-component atmospheric model with the proper equation of state to account for the altered thermodynamics of the sunspot atmosphere due to the formation of \hh.

Our sample was obtained sporadically through the minimum phase of solar cycle 23 and does not contain very large sunspots with high magnetic field strengths.  An intriguing and unexpected finding is the distinctly different behavior of the B$^2$ vs. T curve of the largest sunspots in the survey (e.g., NOAA 11131). While we cannot provide a definitive explanation at this point, we point out that Equation 1 describes only the MHS equilibrium condition for sunspots, and sunspots cannot be in perfect MHS equilibrium all the time (in such case sunspots would be static without possibility of evolution). Therefore we should not expect that the B$^2$ vs. T curves to represent the equilibrium state, or that they follow the same track.  We suspect that differences in the observed behavior of the B$^2$ vs. T curves are the manifestation of changing magnetic and thermal environment of sunspots at different stages of their evolution.  A continued observational effort following sunspots through their lifecycle should provide the necessary data to address this issue as solar cycle 24 enters its maximum phase and larger sunspots start to appear.

\acknowledgments
This work is part of a dissertation submitted to the University of Hawai`i in fulfillment of the requirements for the degree of Doctor of Philosophy.  

The FIRS project was funded by the National Science Foundation Major Research Instrument program, grant number ATM-0421582, and was completed through a collaboration between the Institute for Astronomy and the National Solar Observatory.  We would like to express our profound gratitude to the NSO for all of their assistance and would especially like to thank the DST observers, Doug Gilliam, Joe Elrod, and Mike Bradford for their patience and ingenuity during the commissioning of FIRS.  

Hinode is a Japanese mission developed and launched by ISAS/JAXA, with NAOJ as domestic partner and NASA and STFC (UK) as international partners. It is operated by these agencies in co-operation with ESA and NSC (Norway).  The Hinode SOT/SP Inversions with MERLIN were conducted at NCAR under the framework of the Community Spectro-polarimtetric Analysis Center (CSAC; \url{http://www.csac.hao.ucar.edu/}).  

We would also like to thank the Vector Magnetogram Comparison Group and Andres Asensio Ramos for their essential advice on inversion techniques.  Finally, we would like to thank Huw Morgan, Ali Tritschler, and our referee who have contributed many useful suggestions for improving the text of this article.



\begin{thebibliography}{43}
\expandafter\ifx\csname natexlab\endcsname\relax\def\natexlab#1{#1}\fi

\bibitem[{{Alfv\'en}(1943)}]{alfven43}
{Alfv\'en}, H. 1943, Arkiv f. Math., Astron. o. Fys., 29, 1

\bibitem[{{Balthasar} \& {Schmidt}(1993)}]{balthasar93}
{Balthasar}, H. \& {Schmidt}, W. 1993, A \& A, 279, 243

\bibitem[{{Bartoe} {et~al.}(1979){Bartoe}, {Brueckner}, \& {Jordan}}]{bartoe79}
{Bartoe}, J.-D.~F., {Brueckner}, G.~E., \& {Jordan}, C. 1979, MNRAS, 187, 463

\bibitem[{{Berdyugina} \& {Solanki}(2002)}]{berdyugina02}
{Berdyugina}, S.~V. \& {Solanki}, S.~K. 2002, A \& A, 385, 701

\bibitem[{{Biermann}(1941)}]{biermann41}
{Biermann}, L. 1941, Astron. Ges., 76, 248

\bibitem[{{Borrero} {et~al.}(2003){Borrero}, {Bellot Rubio}, {Barklem}, \& {del
  Toro Iniesta}}]{borrero03}
{Borrero}, J.~M., {Bellot Rubio}, L.~R., {Barklem}, P.~S., \& {del Toro
  Iniesta}, J.~C. 2003, A \& A, 404, 749

\bibitem[{{Cowling}(1976)}]{cowling76}
{Cowling}, T.~G. 1976, Magnetohydrodynamics, Monographs on Astronomical
  Subjects (Hilger)

\bibitem[{{Deinzer}(1965)}]{deinzer65}
{Deinzer}, W. 1965, ApJ, 141, 548

\bibitem[{{Grevesse} \& {Sauval}(1994)}]{grevesse94}
{Grevesse}, N. \& {Sauval}, A.~J. 1994, in Lecture Notes in Physics, Vol. 428,
  Molecules in the Stellar Environment, ed. U.~G. {J\/{o}rgensen}
  (Springer-Verlag), 196

\bibitem[{{Gurman} \& {House}(1981)}]{gurman81}
{Gurman}, J.~B. \& {House}, L.~L. 1981, Sol. Phys., 71, 5

\bibitem[{{Hale}(1908)}]{hale08}
{Hale}, G.~E. 1908, ApJ, 28, 315

\bibitem[{{Hauschildt} {et~al.}(1997){Hauschildt}, {Baron}, \&
  {Allard}}]{hauschildt97}
{Hauschildt}, P.~H., {Baron}, E., \& {Allard}, F. 1997, ApJ, 483, 390

\bibitem[{{Ichimoto} {et~al.}(2008){Ichimoto}, {Lites}, {Elmore}, {Suematsu},
  {Tsuneta}, {Katsukawa}, {Shimizu}, {Shine}, {Tarbell}, {Title}, {Kiyohara},
  {Shinoda}, {Card}, {Lecinski}, {Streander}, {Nakagiri}, {Miyashita},
  {Noguchi}, {Hoffmann}, \& {Cruz}}]{ichimoto08}
{Ichimoto}, K., {Lites}, B., {Elmore}, D., {Suematsu}, Y., {Tsuneta}, S.,
  {Katsukawa}, Y., {Shimizu}, T., {Shine}, R., {Tarbell}, T., {Title}, A.,
  {Kiyohara}, J., {Shinoda}, K., {Card}, G., {Lecinski}, A., {Streander}, K.,
  {Nakagiri}, M., {Miyashita}, M., {Noguchi}, M., {Hoffmann}, C., \& {Cruz}, T.
  2008, Sol. Phys., 248, 233

\bibitem[{{Innes}(2008)}]{innes08}
{Innes}, D.~E. 2008, A \& A L, 481, 41

\bibitem[{{Jaeggli}(2011)}]{jaeggliphdt}
{Jaeggli}, S.~A. 2011, PhD thesis, University of Hawai'i

\bibitem[{{Jaeggli} {et~al.}(2010){Jaeggli}, {Lin}, {Mickey}, {Kuhn}, {Hegwer},
  {Rimmele, T.~R.}, \& {Penn}}]{jaeggli10}
{Jaeggli}, S.~A., {Lin}, H., {Mickey}, D.~L., {Kuhn}, J.~R., {Hegwer}, S.~L.,
  {Rimmele, T.~R.}, \& {Penn}, M.~J. 2010, in MmSAI, Vol.~81, Chromospheric
  Structure and Dynamics: From Old Wisdom to New Insights, 763

\bibitem[{{Jefferies} {et~al.}(1989){Jefferies}, {Lites}, \&
  {Skumanich}}]{jefferies89}
{Jefferies}, J., {Lites}, B.~W., \& {Skumanich}, A. 1989, ApJ, 343, 920

\bibitem[{{Jordan} {et~al.}(1978){Jordan}, {Brueckner}, {Bartoe}, {Sandlin}, \&
  {VanHoosier}}]{jordan78}
{Jordan}, C., {Brueckner}, G.~E., {Bartoe}, J.-D.~F., {Sandlin}, G.~D., \&
  {VanHoosier}, M.~E. 1978, ApJ, 226, 687

\bibitem[{{Kopp} \& {Rabin}(1993)}]{kopp93}
{Kopp}, G. \& {Rabin}, D. 1993, ApJ, 69, 69

\bibitem[{{Kuhn} {et~al.}(1994){Kuhn}, {Balasubramaniam}, {Kopp}, {Penn},
  {Dombard}, \& {Lin}}]{kuhn94}
{Kuhn}, J.~R., {Balasubramaniam}, K.~S., {Kopp}, G., {Penn}, M.~J., {Dombard},
  A.~J., \& {Lin}, H. 1994, Sol. Phys., 153, 143

\bibitem[{{Lin}(1995)}]{lin95}
{Lin}, H. 1995, ApJ, 446, 421

\bibitem[{Lin {et~al.}(1998)Lin, Penn, \& Kuhn}]{lin98}
Lin, H., Penn, M.~J., \& Kuhn, J.~R. 1998, ApJ, 493, 978

\bibitem[{{Lites} {et~al.}(2007){Lites}, {Casini}, {Garcia}, \&
  {Socas-Navarro}}]{lites07}
{Lites}, B.~W., {Casini}, R., {Garcia}, J., \& {Socas-Navarro}, H. 2007, in
  MmSAI, Vol.~78, Solar Magnetism and Dynamics and Themis User Meeting, 148

\bibitem[{Lites {et~al.}(1993)Lites, Elmore, Seagraves, \& Skumanich}]{lites93}
Lites, B.~W., Elmore, D.~F., Seagraves, P., \& Skumanich, A.~P. 1993, ApJ, 418,
  928

\bibitem[{Livingston(2002)}]{livingston02}
Livingston, W. 2002, Sol. Phys., 207, 41

\bibitem[{Maltby {et~al.}(1986)Maltby, Avrett, Carlsson, Kjeldseth-Moe, Kurucz,
  \& Loeser}]{maltby86}
Maltby, P., Avrett, E.~H., Carlsson, M., Kjeldseth-Moe, O., Kurucz, R.~L., \&
  Loeser, R. 1986, ApJ, 306, 284

\bibitem[{Mart{\'\i}nez~Pillet \& V{\'a}zquez(1993)}]{martinez93}
Mart{\'\i}nez~Pillet, V. \& V{\'a}zquez, M. 1993, A \& A, 270, 494

\bibitem[{Mathew {et~al.}(2004)Mathew, Solanki, Lagg, Collados, M., \&
  Berdyugina}]{mathew04}
Mathew, S.~K., Solanki, S.~K., Lagg, A., Collados, M., M., B.~J., \&
  Berdyugina, S. 2004, A \& A, 422, 693

\bibitem[{{Meyer} {et~al.}(1977){Meyer}, {Schmidt}, \& {Weiss}}]{meyer77}
{Meyer}, F., {Schmidt}, H.~U., \& {Weiss}, N.~O. 1977, MNRAS, 179, 741

\bibitem[{Penn {et~al.}(2003{\natexlab{a}})Penn, Cao, Walton, Chapman, \&
  Livingston}]{penn03b}
Penn, M.~J., Cao, W.~D., Walton, S.~R., Chapman, G.~A., \& Livingston, W.
  2003{\natexlab{a}}, Sol. Phys., 215, 87

\bibitem[{Penn {et~al.}(2003{\natexlab{b}})Penn, Ceja, Bell, Frye, \&
  Linck}]{penn03a}
Penn, M.~J., Ceja, J.~A., Bell, E., Frye, G., \& Linck, R. 2003{\natexlab{b}},
  Sol. Phys., 213, 55

\bibitem[{Penn {et~al.}(2002)Penn, Walton, Chapman, Ceja, \& Plick}]{penn02}
Penn, M.~J., Walton, S., Chapman, G., Ceja, J., \& Plick, W. 2002, Sol. Phys.,
  205, 53

\bibitem[{{Rempel}(2010)}]{rempel10}
{Rempel}, M. 2010, in Physics of Sun and Star Spots, IAU Symposium No. 273
  (arXiv:1011.0981v1)

\bibitem[{Rimmele {et~al.}(2004)Rimmele, Richards, Hegwer, Fletcher, Gregory,
  Moretto, Didkovsky, Denker, Dolgushin, Goode, Langlois, Marino, \&
  Marquette}]{rimmele04}
Rimmele, T., Richards, C., Hegwer, S., Fletcher, S., Gregory, S., Moretto, G.,
  Didkovsky, L.~V., Denker, C.~J., Dolgushin, A., Goode, P.~R., Langlois, M.,
  Marino, J., \& Marquette, W. 2004, SPIE, 5171, 179

\bibitem[{{Sanchez Almeida} {et~al.}(1996){Sanchez Almeida}, {Ruiz Cobo}, \&
  {del Toro Iniesta}}]{sanchez96}
{Sanchez Almeida}, J., {Ruiz Cobo}, B., \& {del Toro Iniesta}, J.~C. 1996, A \&
  A, 314, 295

\bibitem[{{Skumanich} \& {Lites}(1987)}]{skumanich87}
{Skumanich}, A. \& {Lites}, B.~W. 1987, ApJ, 322, 473

\bibitem[{Solanki {et~al.}(1992)Solanki, Ruedi, \& Livingston}]{solanki92}
Solanki, S.~K., Ruedi, I., \& Livingston, W. 1992, A \& A, 263, 312

\bibitem[{Solanki {et~al.}(1993)Solanki, Walther, \& Livingston}]{solanki93}
Solanki, S.~K., Walther, U., \& Livingston, W. 1993, A \& A, 277, 639

\bibitem[{Stanchfield {et~al.}(1997)Stanchfield, Thomas, \&
  Lites}]{stanchfield97}
Stanchfield, D.~C., Thomas, J.~H., \& Lites, B.~W. 1997, ApJ, 477, 485

\bibitem[{{Uitenbroek}(2000{\natexlab{a}})}]{uitenbroek00a}
{Uitenbroek}, H. 2000{\natexlab{a}}, ApJ, 531, 571

\bibitem[{{Uitenbroek}(2000{\natexlab{b}})}]{uitenbroek00b}
---. 2000{\natexlab{b}}, ApJ, 536, 481

\bibitem[{{Wallace} {et~al.}(2001){Wallace}, {Hinkle}, \&
  {Livingston}}]{wallace01}
{Wallace}, L., {Hinkle}, K., \& {Livingston}, W.~C. 2001, Sunspot unbral
  spectra in the region 4000 to 8640 cm(-1) (1.16 to 2.50 [microns]) (NSO
  Technical Report)

\bibitem[{Westendorp~Plaza {et~al.}(2001)Westendorp~Plaza, Del Toro~Iniesta,
  Ruiz~Cobo, Mart{\'\i}nez~Pillet, Lites, \& Skumanich}]{westendorp01}
Westendorp~Plaza, C., Del Toro~Iniesta, J.~C., Ruiz~Cobo, B.,
  Mart{\'\i}nez~Pillet, V., Lites, B.~W., \& Skumanich, A. 2001, ApJ, 547, 1130

\end{thebibliography}
\end{document}